\documentclass[10pt]{article}
\pdfoutput=1
\usepackage[utf8]{inputenc}
\usepackage{cite}
\usepackage{amsmath,amssymb,amsbsy,amstext,amsthm,simplewick,amsfonts}
\usepackage{graphicx}
\usepackage{subcaption}
\usepackage{lscape}
\usepackage{cancel}
\usepackage{wrapfig}
\usepackage{upgreek}
\usepackage{bm} %--> bold math...
\usepackage{framed}
\usepackage{bbm}
\usepackage{textcomp}
\usepackage{tikz}
\usepackage{pifont}
\usetikzlibrary{matrix,shapes,fit,tikzmark,calc}
\usepackage{adjustbox}
\usepackage{makecell}
\usepackage{tcolorbox}
\usepackage{physics}
\usepackage{empheq}
\usepackage[normalem]{ulem}
\usepackage{enumitem}
\usepackage{braket} 
\usepackage{array}
\usepackage{dsfont}
\usepackage{ulem}
\usepackage{multirow}
\usepackage{titlesec}
\titleformat{\section}{\normalfont\fontsize{12}{16}\bfseries}{\thesection}{1em}{}

\numberwithin{equation}{section}

\def\a{\alpha}

\def\be{\begin{equation}}
\def\ee{\end{equation}}

\def\d{\delta}

\def\f{\phi}

\def\ba{\begin{eqnarray}}
\def\ea{\end{eqnarray}}

\def\L{\mathcal{L}}

\def\O{\Omega}
\def\bfx{\textbf{x}}

\def\bfk{\textbf{k}}

\def\bfx{\textbf{x}}

\def\del{\partial}

\definecolor{blue3}{RGB}{31,119,180}
\definecolor{red3}{RGB}{214,39,40}
\definecolor{orange3}{RGB}{255,127,14}
\definecolor{green3}{RGB}{44,160,44}

\definecolor{greyish2}{rgb}{.96,.96,.96}

\newcommand{\then}{\quad\Rightarrow\quad}

\usepackage{colortbl}
\definecolor{lightgreen}{cmyk}{0.2, 0, 0.2, 0.2}
\definecolor{lightgray}{cmyk}{0.1,0.2,0,0.1}
\definecolor{lightgray2}{cmyk}{0.1,0.1,0,0.1}

\usepackage{hyperref}
\hypersetup{colorlinks=true,linkcolor=teal,citecolor=orange3,urlcolor=green3,pdfencoding=auto}

\makeatletter
\newlength{\apb@width}
\newcommand{\autoparbox}[2][c]{\settowidth{\apb@width}{#2}\parbox[#1]{\apb@width}{#2}}

\makeatother


\def\d{{\rm d}}

\def\O{{\cal O}}

\def\Z{{\cal Z}}

\def\beq{\begin{equation}}
\def\eeq{\end{equation}}

\newcommand{\dt}{\tilde \delta_{D}^{(3)}}

\newcommand{\vk}{\mathbf{k}}

\newcommand{\la}{\lambda}
\newcommand{\rd}{{\rm d}}

\allowdisplaybreaks[1]
\begin{document}

%\newgeometry{top=2cm, bottom=2cm, left=2.9cm, right=2.9cm}

\begin{titlepage}
\setcounter{page}{1} \baselineskip=15.5pt

\thispagestyle{empty}

\renewcommand*{\thefootnote}{\fnsymbol{footnote}}

\begin{center}

{\fontsize
{20}{20} \bf A Cosmological Bootstrap\\[0.5cm] for Resonant Non-Gaussianity} \\

% {\fontsize
% {20}{20} \bf What is the bispectrum\\[0.5cm] for a complex number of derivatives?} \\

\end{center}

\vskip 18pt
\begin{center}
\noindent
{\fontsize{12}{18}\selectfont Carlos Duaso Pueyo\footnote{\tt cd820@cam.ac.uk} and Enrico Pajer\footnote{\tt enrico.pajer@gmail.com}}
\end{center}

\begin{center}
\vskip 8pt
\textit{Department of Applied Mathematics and Theoretical Physics, University of Cambridge, \\
Wilberforce Road, Cambridge, CB3 0WA, UK} 
\end{center}

%=========================================

\vspace{2cm}

%The answer to the title is resonant non-Gaussianity :)
\noindent 

Recent progress has revealed a number of constraints that cosmological correlators and the closely related field-theoretic wavefunction must obey as a consequence of unitarity, locality, causality and the choice of initial state. When combined with symmetries, namely homogeneity, isotropy and scale invariance, these constraints enable one to compute large classes of simple observables, an approach known as (boostless) cosmological bootstrap. Here we show that it is possible to relax the restriction of scale invariance, if one retains a discrete scaling subgroup. We find an infinite class of solutions to the weaker bootstrap constraints and show that they reproduce and extend resonant non-Gaussianity, which arises in well-motivated models such as axion monodromy inflation. We find no evidence of the new non-Gaussian shapes in the Planck data. Intriguingly, our results can be re-interpreted as a deformation of the scale-invariant case to include a complex order of the total energy pole, or more evocatively interactions with a complex number of derivatives. We also discuss for the first time IR-divergent resonant contributions and highlight an inconsequential inconsistency in the previous literature.

%=========================================

\end{titlepage}

\setcounter{tocdepth}{2}
{
\hypersetup{linkcolor=black}
\tableofcontents
}

\renewcommand*{\thefootnote}{\arabic{footnote}}
\setcounter{footnote}{0} 

\newpage

%%%%%%%%%%%%%%%%%%%%%%%%%%%%%%%%%%%%%%%%%%%%%%%%%%%%%%%%%%%%%%%%%%
\section{Introduction}

In the study of a physical system, much of the initial effort is devoted to charting the landscape of theories that are both consistent and compatible with the data. In this endeavor, we can either sample the bulk of theories by building concrete models, or probe the boundaries of theory space using fundamental principles of physics. These two approaches, which can be roughly labelled as model building and bootstrapping, complement each other in very powerful ways and particle physics is a spectacular success story of this synergy in action. Much recent work aspires to achieve similar progress in the study of the primordial universe and the origin of initial conditions for our universe \cite{Baumann:2022jpr}. \\
Both model building and bootstrapping heavily leverage symmetries to reduce the complexity of the analysis and to sharpen physical predictions. For models of the early universe, a major role has been played by scale invariance, which in the cosmological context\footnote{In the language of conformal field theories what cosmologists call ``scale invariance'' is the scaling property of an operator of zero conformal dimension.} refers to a fixed scaling behaviour of $n$-point primordial correlation functions of curvature perturbations, $B_n(\lambda \bfk ) \sim \lambda^{3(1-n)} B_n(\bfk)$. Scale invariance has been observed to be an approximate symmetry of primordial perturbations, with percent deviations detected in the so-called spectral tilt of the scalar power spectrum. While it is natural to focus on models that explain such a surprising and unexpected symmetry, such as slow-roll inflation, it is also interesting to explore possible deviations from it that are still allowed by the data. At the phenomenological level, a minimal possibility is to reduce \textit{continue} scale invariance to a \textit{discrete} subgroup, by allowing for small periodic oscillations of correlators with scale, which average out over a large range of scales (for early proposals see e.g. \cite{Chen:2008wn}). Remarkably, this possibility appears to be also natural from the model building point of view, where the same setup that ensures continue scale invariance in perturbation theory displays small periodic deviations from it due to non-perturbative effects \cite{Freese:1990rb,Silverstein:2008sg,McAllister:2008hb,Flauger:2009ab,Kaloper:2011jz,Pajer:2013fsa}.\\

The goal of this work is to study cosmological correlators that enjoy only discrete scale invariance, as in models of resonant non-Gaussianity \cite{Chen:2008wn,Flauger:2010ja}. We perform this analysis by adapting the tools of the boostless cosmological bootstrap, recently developed in a series of papers \cite{BBBB,COT,MLT,sCOTt,Goodhew:2021oqg}, building upon previous work \cite{Arkani-Hamed:2017fdk,Arkani-Hamed:2018kmz,Green:2020ebl}, to the case where full discrete scale invariance is not a symmetry. Our analysis and results can be motivated by several points of views.

\paragraph{Resonant non-Gaussianity} At the phenomenological level, we derive an infinite class of ``resonant" scalar bispectra, which we then compare to CMB data by Planck to obtain observational constraints. The leading resonant non-Gaussianity had already been studied in a concrete model in \cite{Flauger:2010ja} and within the framework of the effective field theory of inflation \cite{Cheung:2007st} in \cite{Behbahani:2011it} (see also \cite{Gwyn:2012pb,Meerburg:2010ks,Leblond:2010yq,Abolhasani:2019lwu,Abolhasani:2020xcg} for other studies). Here we find infinitely many new shapes that correspond to cubic scalar operators with higher derivative interactions. In a technically natural effective theory, higher-dimension operators are generically expected to produce smaller signals. While the calculation can be performed exactly, the regime of fast oscillations is particularly interesting because the result simplifies considerably and the signal is expected to be relatively large. This regime is controlled by the dimensionless parameter $\alpha$, namely the frequency $\omega$ of background oscillations that are responsible for the resonance in units of the Hubble parameter, $\alpha = \omega/H$. For the bispectra that had already been discussed in the literature, our derivation reproduces the same leading order term in $\alpha \gg 1$. However we find a different result at the next-to-leading order in $1/\alpha$. Within our derivation the discrepancy is unavoidable because the resonant bispectra published in the literature do not satisfy the so-called manifestly local test, a necessary condition for the non-Gaussianity generated by local interactions \cite{MLT}. Upon further inspection we identify the source of the discrepancy: in previous calculations the stationary phase approximation was employed, but subleading corrections in $\alpha$ had not been correctly propagated to the final result, hence yielding an incorrect prediction of the next to leading and subsequent terms. Amending the mistake in the literature is in practice inconsequential for phenomenology, since it amounts to a small shift in the phase of oscillations, which is a free parameter of the model. However, the situation is a good reminder of the usefulness of model-independent constraints, both for bootstrapping and for checking the consistency of predictions.\\

\paragraph{The bootstrap without scale invariance} There is a second independent motivation for our analysis. So far all cosmological bootstrap approaches have made crucial use of scale invariance. Indeed, the de Sitter invariant approach of \cite{Arkani-Hamed:2018kmz,Baumann:2019oyu,Baumann:2020dch,Hillman:2019wgh} relied on the full group of dS isometries and the same applies to results obtained from ``momentum-Mellin" space \cite{Sleight:2019mgd,Sleight:2019hfp} or Anti-de Sitter \cite{Sleight:2021plv,Sleight:2020obc}. In a second approach, motivated by the no-go theorems of \cite{Green:2020ebl}, invariance under dS boosts is relaxed \cite{BBBB}, and additional constraints from locality \cite{MLT}, unitarity \cite{COT,Cespedes:2020xqq,Goodhew:2021oqg,Albayrak:2023hie}, and causality/analyticity \cite{Arkani-Hamed:2018bjr,Arkani-Hamed:2018kmz,Salcedo2022,Agui-Salcedo:2023wlq,Ghosh:2022cny} are brought into the game, while still assuming scale invariance (see e.g. \cite{Bonifacio:2021azc,Bonifacio:2022vwa,Baumann:2021fxj,Hillman:2021bnk,Cabass:2021fnw,Cabass:2022jda,Ghosh:2023agt}). Several hybrid approaches have leveraged the dS invariance of a seed signal and combined it with the breaking of boosts in a systematic way (see e.g. \cite{Pimentel:2022fsc,Jazayeri:2022kjy,Wang:2022eop,Jazayeri:2023xcj}). It is therefore natural to ask how far one can go when scale invariance is relaxed. Here we show that the bootstrap approach can be successful in predicting a class of new non-Gaussian signals, even when the constraint of scale invariance has been weakened to a discrete subset. The price to pay is that we will have to \textit{rely more heavily on the structure of the initial ansatz}. At a hand-wavy level, the ansatz can be justified by invoking the ``simplest possible analytic structure'', however we don't attempt to formalise this in a precise mathematical sense. An interesting byproduct of our analysis is that we highlight interesting modifications of well-known bootstrap constraints. For example, the Cosmological Optical Theorem (COT) does apply to the case of resonant non-Gaussianity (and indeed to any FLRW spacetime with time-dependent couplings \cite{Goodhew:2021oqg}). However this constraint relates leading terms to exponentially suppressed terms and is therefore not very useful unless one attempts to bootstrap the full result. Instead, we show that the COT can be modified by taking advantage of the analytic dependence of the result on a non-kinematical parameter, namely the frequency of oscillations $\alpha$. This leads to a \textit{modified cosmological optical theorem} that constrains directly the leading non-Gaussianity, which in practice allows us to completely neglect the inconsequential exponentially suppressed terms. \\

\paragraph{A complex deformation} Finally, we come to the third and last motivation. At face value, cosmological correlators are functions of the kinematics and of parameters in the theory. The kinematical dependence has been explored extensively, while the parameter dependence appears to be quite trivial in perturbation theory: each contribution is simply proportional to the product of the coupling constants involved in the corresponding diagram. However, there is one parameter that appears non-trivially. To see this, consider a contact interaction of a massless scalars in dS. At tree-level one expects a contribution to the wavefunction coefficients of the form
\begin{align}
    \psi_n(\{\bfk\})\sim \frac{\text{Poly}(\{\bfk\})}{k_T^p}\,,
\end{align}
where $\{\bfk\}$ collectively denotes the $n$ three-momenta $\bfk_a$ and the so-called total energy is the sum of their norms, $k_t\equiv \sum_a |\bfk_a|$. Assuming scale invariance, it was shown that $p$ is a positive integer\footnote{The special case of $p=0$ can correspond to a logarithmic IR divergence in perturbation theory that is associated to interactions with few or no derivatives. Such contributions are severely constrained by unitarity \cite{COT} and are crucial in the context of parity violating interactions \cite{Cabass:2021fnw,Cabass:2022rhr,Stefanyszyn:2023qov,Creque-Sarbinowski:2023wmb}. We will discuss IR divergent resonant contributions in Section \ref{ss:IR}} given by \cite{BBBB}
\begin{align}
    p=1+\sum_A (\Delta_A-4)\,,
\end{align}
where the sum runs over each interaction of a given diagram and $\Delta_A$ is the mass dimension of the interaction at vertex $A$. This result applies also to the order of partial energy singularities and assumes the absence of IR divergences. For example, for $n=3$ we need a cubic interaction and so $p$ simply counts the total number of space and time derivatives. Surprisingly, the interesting parameter for us here is $p$. Indeed, the results in this paper show that resonant non-Gaussianity can be roughly thought of as a continuation to complex $p$ of the standard scale-invariant result (albeit with different constraints on the polynomial). Indeed, taking $p\to p+i\alpha$, we obtain non-Gaussianity of the form
\begin{align}\label{complexdef}
    \psi_n(\{\bfk\})\sim \frac{\text{Poly}(\{\bfk\})}{k_T^{p+i\alpha}}\sim \frac{\text{Poly}(\{\bfk\})}{k_T^{p}} e^{-i\alpha \log(k_T)}  \then  B_n(\{\bfk\})\sim  \frac{\text{Poly}(\{\bfk\})}{k_T^{p}} \cos(\alpha \log(k_T)) \,,
\end{align}
where the scale that makes the logarithm well-defined corresponds to a model-dependent shift in the phase of oscillations, which is marginalized over when comparing to observations. From the perspective of the bulk time integral representation, adding an imaginary part to $p$ corresponds to a complex twist of the integral
\begin{align}
    \int \frac{d\eta}{\eta} \, \eta^{p} e^{ik_T\eta}\to \int \frac{d\eta}{\eta} \, \eta^{p+i \a} e^{ik_T\eta}\,,
\end{align}
where $\omega = \alpha H$ and $\exp(i\alpha \log \eta) \sim \cos(\omega t)$ represents a background oscillation that is periodic in cosmological time $t$. This interpretation is quite close in spirit to the proposal of \cite{Arkani-Hamed:2017fdk} of relating what ultimately are flat space objects to cosmological wavefunctions by integrating over appropriate energy shifts of the interaction vertices. Indeed, resonant non-Gaussianity is produced on sub-Hubble scales, around $k\eta \sim \alpha \gg 1$, where the flat-space approximation becomes increasingly good. As a consequence, the \textit{full} correlator is specified by a corresponding flat-space amplitude $A_n$, to leading order in $\alpha$,
\begin{align}
    B_n \sim \frac{A_n}{k_T^p (\prod_a^n k_a)} \cos(\alpha \log(k_T)) \left[1+ \mathcal{O}(1/\alpha)\right]\,.
\end{align}
It's important to contrast this with the usual statement that the UV-limit of a flat-space amplitude appears as the residue of the leading $k_T$ pole of a wavefunction coefficient \cite{Raju:2012zr,Maldacena:2011nz}. In the usual case the leading $k_T\to 0$ pole is leading only for unphysical complex momenta, while for physical momenta many other contributions arise, namely all subleading $k_T$ poles, that are of the same order as the leading $k_T$-pole \textit{and} are not directly related to flat space. In contrast, for resonant non-Gaussianity we have a physical parameter, namely $\alpha$, that we can dial to make the flat-space amplitude contribution leading for all\footnote{The order of limits is important here and we are assuming that we first take $ \alpha \gg 1$ and then choose a certain kinematical configuration. Indeed, by pushing deep into a certain kinematical configurations, such as squeezed and collinear limits, while keeping other parameters fixed, we always probe physics being the low-energy effective theory \cite{Arkani-Hamed:2015bza}.} physical momenta.

%%%%%%%%%%%%%%%%%%%%%%%%%%%%%%%%%%%%%%%%%%%%%%%%%%%%%%%%%%

\paragraph{Summary of the results} For the convenience of the busy modern reader, we summarize here the take-home messages of this work:
\begin{itemize}
    \item We have developed a (boostless) cosmological bootstrap that does not rely on scale invariance but only on a weaker discrete scaling subgroup, determined by the condition~\eqref{eq:avgsi}. We find an infinite class of non-Gaussianities (cubic wavefunction coefficients and bispectra) that satisfy this relaxed scaling condition as well as all other known physical constraints imposed by unitarity in the form of the cosmological optical theorem, locality in the form of the manifestly local test, and the choice of initial state in the form of the sufficiently constraining ansatz in \eqref{simply}. The result, summarized to all orders in derivatives in \eqref{eq:finalres}, turns out to be a generalization of the well-known resonant non-Gaussianity \cite{Flauger:2010ja}, which arises in models of axion monodromy inflation. Our results are possibly not exhaustive of all possibilities with the chosen symmetry, but are expected to capture the leading signal. We comment in Section \ref{sec:c-} on additional contributions from the resonant excitation of the mode functions, which we neglect in the main derivation. 
    \item As discussed above, our results are close to the scale invariant case and can be thought of as a deformation of the order $p$ of the total energy pole to include an imaginary part, which in turn leads to oscillations in the total energy $k_T$, and breaks scale invariance down to a discrete subgroup, as in \eqref{complexdef}.
    \item For the first time we discuss IR divergent resonant contributions (see Section \ref{ss:IR}). These appear in the wavefunction coefficients as well as in the correlators of a spectator scalar. However they disappear in the bispectrum of curvature perturbations in single field inflation, as expected on general grounds.
    \item We searched for several resonant non-Gaussian bispectra in the temperature anisotropies of the cosmic microwave background, as measured by the Planck satellite using the pipeline CMB-BEST \cite{Sohn2023cmbbest}. We find no evidence of a signal, as discussed in Section \ref{sec:cmb}. 
    \item Even though the cosmological optical theorem (COT) applies to time dependent couplings, it relates leading terms, which we bootstrap, to exponentially suppressed terms, which we wish to neglect and so is not practically useful. However, we derive a modified cosmological optical theorem that can be applied to the case of an \textit{imaginary} and time-dependent coupling constant $ e^{\pm i \omega t}$. This is useful because the modified COT relates the leading term in the wavefunction coefficient to itself and hence determines the reality of its coefficients. The new relation leverages the simple analytic dependence of the result on a parameter of the model, namely the frequency of oscillations $\omega$ or its dimensionless cousin $\alpha=\omega/H$.
\end{itemize}

We end this introduction with a guide to the organization of the rest of this paper. In Section \ref{sec:rules} we discuss a set of bootstrap rules and put forward a bootstrap ansatz, which will be subsequently used to find an infinite class of resonant non-Gaussianities. In particular, we discuss the constraint from discrete scale invariance, the form of the ansatz appropriate for resonant production of non-gaussianity from a Bunch-Davies initial state, the modified cosmological optical theorem that applies to the leading resonant contributions and finally the manifestly local test. In Section \ref{sec:bootstrap} we use these constraints to derive an infinite class of cubic wavefunction coefficients and present some examples to the lowest orders in derivatives. We also study IR divergences in Subsection \ref{ss:IR} and parity-odd interactions in Subsection \ref{ss:PO}. Then, in Section \ref{sec:c-}, we discuss subleading contributions coming from the resonant excitation of negative frequency modes in the free theory and their impact on non-Gaussianity. In Section \ref{sec:cmb}, we present the observational constraints on the leading resonant bispectra that we have derived from the Planck data on CMB temperature and polarization anisotropies. We conclude in Section \ref{sec:conclusions} with a discussion and an outlook.

%%%%%%%%%%%%%%%%%%%%%%%%%%%%%%%%%%%%%%%%%%%%%%%%%%%%%%%%%%%%%%%%%%
\section{Bootstrap ansatz and rules} \label{sec:rules}

In this section, we present a set of bootstrap rules and a bootstrap ansatz. Then we show how they lead to a class of solutions for the cubic wavefunction coefficient of a massless scalar field in de Sitter spacetime. Our derivation parallels that in \cite{BBBB,MLT}, but we relax the requirement of exact continuous scale invariance in favour of discrete scale invariance.

Anticipating invariance under translations, we parameterize the field-theoretic wavefunction as
\begin{align}\label{psin}
\Psi[\phi;\eta]=\exp\left[   +\sum_{n}^{\infty}\frac{1}{n!} \int_{\bfk_{1},\dots,\bfk_{n}}\,  \dt  \left( \sum_{a}^{n} \bfk_{a} \right) \psi_{n}(\{\bfk\};\eta)  \phi(\bfk_{1})\dots \phi(\bfk_{n})\right]\,,
\end{align}
where to simplify the notation we have introduced %the rescaled Dirac delta
\begin{align}
\int_{\bfk} \equiv \int \frac{\rd^3 k}{(2\pi)^{3}} \, , \qquad\qquad \dt(\bfk)\equiv (2\pi)^{3} \delta_D^{(3)}\left(  \bfk \right)\,,
\end{align}
and $\eta$ is conformal time for the flat (Poincar\'e) patch of de Sitter spacetime,
\begin{align}\label{ds}
    ds^2=\frac{-d\eta^2+d\bfx^i\d\bfx^j \delta_{ij}}{\eta^2 H^2}\,.
\end{align}
Throughout we assume that $\phi$ is a spectator massless scalar field in de Sitter and we allow for boost-breaking interactions.

%%%%%%%%%%%%%%%%%%%%
\subsection{Homogeneity and isotropy}

An $n$-point wavefunction coefficient $\psi_n (\{\vk\};\eta)$ is in general a function of the $n$ wavenumbers $\vk_a$ with $a=1,\dots,n$ and of time $\eta$. We will focus on the late time behaviour of $\psi_3$, as $\eta \to 0$, namely at the future conformal boundary of dS.

In writing \eqref{psin} we already assumed invariance under spatial translations, which leads to the momentum conserving delta function. We furthermore impose invariance under rotations, which implies that $\psi_n$ is a function of rotation-invariant contractions of the momenta with the Kronecker delta $\delta_{ij}$ or the Levi-Civita symbol $e_{ijk}$. Invariance under rotations and translations reduces the number of variables on which $\psi_n$ depends from $3n$ to $3(n-2)$. Here we will focus on $n=3$. In this case the Levi-Civita symbol cannot appear because all the momenta lie in the same plane. Hence $\psi_3$ is a function of just three variables, which can be chosen to be the norms $k_a\equiv |\vk_a|$ of the momenta
\beq
	\psi_3 (\vk_1,\vk_2,\vk_3) =  \psi_3 (k_1,k_2,k_3) \, .
\eeq
Let us discuss the constraints on $ \psi_3 (k_1,k_2,k_3)$ from a particular form of discrete scale invariance.

%%%%%%%%%%%%%%%%%%%%
\subsection{Discrete scale invariance}

Usually one is interested in imposing continuous scale invariance, namely the relation %\enr{can $\lambda$ really be negative?}
\beq\label{scaleinv}
	\psi_3 (\lambda k_a) = \lambda^3 \, \psi_3 (k_a) \qquad \forall \,\, \lambda\in\mathbb{R}^+ \, .
\eeq
This transformation is the manifestation of the dilation isometry of dS for massless fields at the future conformal boundary, $\eta \to 0$. To see this, notice that a dilation rescales $\bfx \to \lambda \bfx$ and $\eta \to\lambda \eta$, which leaves the line element in \eqref{ds} invariant. For a massless scalar we expect that the leading time dependence for $\eta \to 0 $ is a constant, $\phi \sim \eta^0$. For the late-time wavefunction this in turn leads to~\eqref{scaleinv}.\footnote{More generally, for a field of mass $m$ one finds the leading late-time scaling $\phi \sim \eta^\Delta $ with $\Delta=3/2+\sqrt{9/4-m^2/H^2}$. Then the wavefunction coefficients scale as $\psi_n\sim k^{3+n(\Delta-3)}$.}

Here we are interested in relaxing this condition. If we made no assumption whatsoever about scale invariance, the family of possible solutions would be too large to make any useful predictions. Instead, we will assume that the continuous scale invariance in \eqref{scaleinv} is relaxed to a specific discrete form of scale invariance defined as follows:
\beq
	\exists \quad \bar{\lambda}\in\mathbb{R}^+\setminus \{ 1 \} \qquad\text{s.t.}\qquad \psi_3(\bar{\la}k_a) = \bar{\la}^3 \, \psi_3 (k_a) \, ,
\eeq
where $\bar \lambda$ is some real and positive number which is part of the data that specifies the theory. Notice that this then implies
\beq
	\psi_3(\bar{\la}^n k_a) = \bar{\la}^{3n} \, \psi_3 (k_a) \qquad \forall \,\, n\in\mathbb{Z} \, ,
\eeq
and consequently we can take $\bar{\la}\in (1,\infty)$ without loss of generality. With this information, we finally formulate our \textit{discrete scale invariance} constraint as
\beq \label{eq:avgsi}
	\exists \quad \bar{\lambda}\in (1,\infty) \qquad\text{s.t.}\qquad \psi_3(\bar{\la}^nk_a) = \bar{\la}^{3n} \, \psi_3 (k_a) \qquad \forall \,\, n\in\mathbb{Z} \, .
\eeq
This condition is inspired by a class of models in which periodic oscillations appear in the inflaton potential, 
\begin{align}
    V(\phi) = V_0(\phi)+\Lambda^4 \cos\left(\frac{\phi}{f}\right)\,,
\end{align}
where $V_0$ is some slow-roll potential, and $\Lambda$ and $f$ are energy scales. This structure is a rather generic feature in models of axion inflation, where the periodic features are induced by non-perturbative effects in a gauge sector to which the axion couples via a shift-symmetric dimension five coupling (sometimes called ``Chern-Simons'' or ``theta-term''). A partial collection of references is \cite{Freese:1990rb,Freese:2004vs,Kim:2004rp,McAllister:2008hb,Pahud:2008ae,Flauger:2009ab}. One can generalize this setup by considering the effective field theory of inflation (EFToI) \cite{Creminelli:2006xe,Cheung:2007st}. In the most general case, the Wilsonian coupling constants appearing in the EFToI are arbitrary functions of time. Then one assumes that when they are approximately constant one recovers exact scale-invariant predictions. To generalize to oscillations in the inflaton potential, one demands that the Wilsonian couplings are not constants but are invariant under specific discrete cosmic time translations, $t \to t+{\rm const}$~\cite{Behbahani:2011it,Behbahani:2012be}. The discrete shifts in time lead to a discrete rescaling in $k$ as follows. First notice that in de Sitter $t\propto \ln(-\eta)$ and so a shift in $t$ corresponds to a rescaling of $\eta$. Second, recall that when the Wilsonian coefficients are constant the theory must be invariant under dilations and so only the combination $k\eta$ can appear. Then $ k \sim \eta \sim e^{-Ht}$ and the discrete scale invariance in \eqref{eq:avgsi} follows. Here we bypass this time evolution picture and simply use the constraint in \eqref{eq:avgsi} as a definition of the class of models we are considering. \\

Before proceeding we should mention that here we are assuming that there are no IR divergences and so we can safely evaluate $\psi_3(\eta)$ as $\eta \to 0$. In the presence of IR divergences, we have to introduce a late-time regulator and discrete scale invariance needs to be formulated in a slightly modified way. In this section we will continue assuming no IR divergences, and we postpone a discussion of those to Section \ref{ss:IR}.

%%%%%%%%%%%%%%%%%%%%%%%%%%%%%%%%%%%%%%%%%%%%%%%%%%%%%%%%%%%%%%

\subsection{The bootstrap ansatz} \label{sec:fact}

Here we want to argue that a minimal ansatz that satisfies the constraint from discrete scale invariance in \eqref{eq:avgsi} is given by
\beq \label{eq:factorized}
	\psi_3 (k_a) = e^{-i\a\log{(k_T/k_*)}} \cdot \hat{\psi}_3 (k_a) \simeq \frac{\hat{\psi}_3 (k_a)}{k_T^{i\alpha}} \,,
\eeq
where 
\begin{align}
    k_T\equiv  k_1+k_2+k_3
\end{align}
is the so-called total energy, $k_*$ is some constant pivot scale, which is degenerate with the normalization of $\hat \psi_3$, and $\hat \psi_3$ is a rational function of the energies $k_a$ that scales as
\begin{align}
\hat{\psi}_3 (\lambda k_a) = \lambda^3 \hat{\psi}_3 (k_a) \qquad \forall \,\, \lambda\in\mathbb{R}^+ \, .
\end{align}
Since the bootstrap procedure cannot fix it, we will omit the overall phase $k_*$ throughout, and will only restore it when presenting the final results in Section~\ref{sec:bootstrap}. If we assume manifestly local interactions and a Bunch-Davies initial state, the only singularity that $\hat \psi_3$ can have for complex energies $k_a$ is at $k_T = 0 $ and hence it can be written as 
\begin{align}\label{rulerational}
    \hat \psi_3(k_a)=\frac{\text{Poly}_{p+3}(k_1,k_2,k_3)}{k_T^p}\,,
\end{align} 
where $p$ is the maximum order of the total-energy singularity, which equals the maximum number of space plus time derivatives one is considering, and the numerator is a homogeneous polynomial of degree $p+3$. Notice that combining \eqref{eq:factorized} and \eqref{rulerational} we find simply
\beq \label{simply}
	\psi_3(k_a)=\frac{\text{Poly}_{p+3}(k_1,k_2,k_3)}{k_T^{p+i\alpha}}\,.
\eeq
This is almost identical to the usual ansatz for the scale invariant bispectrum \cite{BBBB}, except for the crucial difference that the ``order of the $k_T$ pole'' is allowed to have an imaginary part. At a more abstract level, this arises by introducing a complex twist into the usual bulk time integrals, which is closely related to the ongoing work of \cite{Gui}. We take \eqref{simply} to be \textit{part of the data that defines the theory we are studying}. Again, we postpone a discussion of possible IR divergences to Section \ref{ss:IR}.

In the following we will justify this ansatz in two ways. First by arguing that this ansatz possesses the ``simplest'' possible analytic structure. Second, we will see that this ansatz is precisely what is expected from the picture of bulk time evolution.

\paragraph{The simplest analytic structure} An imprecise but general lesson of the bootstrap approach to amplitudes and cosmological correlators \cite{Baumann:2022jpr} is that the growing complexity of bulk perturbation theory manifest itself in the increasingly complex analytic structure of the corresponding ``boundary'' quantities as function of the kinematical variables $\bfk_a$. For example, contact diagrams of massless fields have only total energy poles, exchange diagrams have also partial energy poles and loop diagrams display a collection of branch points (see e.g. \cite{Arkani-Hamed:2017fdk} and \cite{Salcedo2022} for more systematic discussions in Minkowski). Hence we would like to start looking for solutions of the discrete scale invariance constraint that have the simplest possible analytic structure. While it would be nice to do this in a mathematically rigorous way by formally quantifying the ``complexity'' of a given analytic structure, here we proceed heuristically and consider first entire functions, then functions with a finite or infinite number of poles and finally functions with a finite or infinite number of branch points. A slightly more detailed discussion is given in Appendix \ref{app:anstr}. First notice that the only way to have a finite number of poles is to have poles at $k=0$ and $k=\infty$. Poles at any other location must have infinitely many images because of discrete scale invariance. In this simplest case $\psi_n$ is forced to also satisfy continuous scale invariance. This case has already been discussed in the literature, e.g. in \cite{BBBB,MLT}, and so we set it aside. The next simplest possibility is to have an infinite number of poles, which necessarily accumulate at $k=0$ and $k=\infty$. We weren't able to find any concrete function that satisfies this property and so we move on to the next possibility---two branch points at $k=0$ and $k=\infty$ connected by a single branch cut (which in some sense can be thought of as a limiting case of an infinite number of poles). The rich class of discrete scale invariant functions with such an analytic structure is given by $e^{-i\a\log{(k_T/k_*)}}$ times a homogeneous function of degree three, which finally brings us to \eqref{eq:factorized}, the ansatz that we will consider in the rest of this paper.

\paragraph{The bulk time evolution} Here we will the use the representation of the cosmological wavefunction in terms of bulk time integrals to confirm our choice of the ansatz in \eqref{simply}. The bulk integrals that are commonly encountered in resonant models at tree level have the form\footnote{Here for concreteness we focus on $\psi_3$ but the whole discussion applies to \textit{all contact} wavefunction coefficients $\psi_n$ for any $n$ with the only difference that for $n\geq4$ one cannot re-write all vector products $\bfk_a \cdot \bfk_b$ in terms of the norms $k_c$. }~\cite{Chen:2010bka}
\beq \label{eq:ints}
	\psi_3^{\text{contact}} (k_a) = \int_{-\infty}^0 \frac{\rd\eta}{\eta^4} \, \cos(\omega t) e^{i  k_T \eta} \, f(\bfk_a,\eta) \, , %\qquad\text{or}\qquad \int_{-\infty}^0 d\eta \, e^{\pm i\omega t} e^{\pm i c_s k_T \eta} \, f(\eta) \, ,
\eeq
where we are omitting any possible overall factors that might depend on $H$, $\omega$, and the couplings. 
Several comments are in order. In this expression, the first factor comes from the background oscillations contained in the coupling constants, the second factor from the bulk-to-boundary propagators, and the function $f(\bfk_a,\eta)$ contains the non-oscillatory part of the bulk-to-boundary propagators of a massless scalar field $\phi$, its derivatives with respect to time, other polynomial factors of $\eta$ coming from the vertices, and factors of momenta $\bfk_a$ coming from spatial derivatives. In \eqref{eq:ints} we assumed Bunch-Davies mode functions (i.e. only positive frequencies $e^{+ i k \eta}$), but resonant models excite a small component of the negative frequency solution. For the moment we neglect these subleading contributions but we will discuss them in Section~\ref{sec:c-}. Assuming the Bunch-Davies massless mode functions
%\beq \label{eq:pimf}
%	\phi^{+}_k (\eta) = \frac{H}{\sqrt{2k^3}} (1-ik\eta ) \, e^{ik\eta} \, ,
%\eeq
the bulk-to-boundary propagator is
\beq\label{BDsol}
	K(k,\eta) = \frac{\phi^+_k (\eta)}{\phi^+_k (0)} = (1-ik\eta) \, e^{ik\eta} \, ,
\eeq
and the $n$-th time derivative takes the from
\beq \label{eq:Ktimed}
	\eta^n\frac{d^n}{d\eta^n} K(k,\eta) = (i \eta k)^n (1-n-ik\eta) \, e^{ik\eta} \, .
\eeq
Hence the contributions to $f(k_a,\eta)$ from the mode functions are polynomial in $k_a \eta$. If we further restrict to \textit{manifestly local interactions}, namely products of fields and their derivatives at the same spacetime point, also time and space derivatives contribute to $f(k_a,\eta)$ additional monomials in $k_a \eta$ (since $\bfk_a\cdot \bfk_b$ can be re-written in terms of $k_a$). Note that $f$ is scale invariant,
\begin{align}
    f(\lambda k_a,\eta/\lambda)=f(  k_a,\eta )\,,
\end{align}
because every factor of $k$ is accompanied by a factor of $\eta$ both in the massless dS mode functions and in the time and space derivatives, where it arises from the metric that contracts the indices. In summary, we can write
\begin{align} \label{eq:f}
    f(k_a,\eta)=\sum_q a_q \, \text{Poly}_q(k_a) \, \eta^q\,,
\end{align}
where $a_q$ are possibly complex numerical coefficients and $\text{Poly}_q$ is a homogeneous polynomial of degree $q$. In passing, we mention that we do allow for a speed of sound $c_s$ that is different from the speed of light, but we will assume throughout that $c_s$ is constant in time. To the order to which we are working, $c_s$ only appears as an overall factor normalizing the size of non-Gaussianity. Since the overall normalization is anyways model dependent, we simply drop all factors of $c_s$ throughout the calculation. \\

The integral in \eqref{eq:ints} gives a pair of gamma functions for each monomial in $f(k_a,\eta)$. To see this, let us use $Ht = -\log(-H\eta)$, the complex exponential representation of the cosine, and let's introduce the dimensionless frequency of oscillations,
\begin{align}
\a\equiv \frac{\omega }{H}\,.
\end{align}
Then, up to overall factors, we can  write a master integral in the following form (here we set $H=1$, which eventually leads to an inconsequential overall factor)
\begin{align}\label{Iint}
    \mathcal{I}_m(k_T) \equiv \int_{-\infty}^0 d\eta \, \eta^{m-1} \, \left[ (-\eta)^{+i\alpha} +  (-\eta)^{-i\alpha}  \right] \,  e^{i  k_T \eta} %= \int_{-\infty}^0 \frac{d\eta}{\eta} \left(  |\eta|^{m-i\alpha} + |\eta|^{m+i\alpha}  \right) \,  e^{i  k_T \eta} \, ,
\end{align}
for some integer $m$ and where, as usual, $\eta \to -\infty$ is reached with a small positive imaginary part so that the integral converges. Under a Wick rotation, this integral can be related to two Euler gamma functions\footnote{For interactions with too few derivatives one might have $m\leq 0$ and the integral does not converge at $\eta=0$. Then one needs to regulate the upper boundary of integration to a finite time $\eta_0<0$. For single field inflation the effective shift symmetry of curvature perturbations ensures that there are always enough derivatives in the interaction so that no IR divergences arise and one can safely take $\eta_0 \to 0$. We will ellaborate on this in Section~\ref{ss:IR}.} with a complex argument
\beq \label{eq:masterintres}
    \mathcal{I}_m (k_T) \simeq - \, e^{\pi\a/2} \frac{i^m}{k_T^{m+i\a}}  \Gamma(m+i\alpha) - e^{-\pi\a/2} \frac{i^m}{k_T^{m-i\a}}  \Gamma(m-i\alpha) \, ,
\eeq
where the factors of $e^{\pm \pi\a /2}$ result from the Wick rotation. This, and the fact that in the large-$\a$ limit both gamma functions behave as $\Gamma (m\pm i\a) \sim e^{-\pi\a/2}$, implies a relative exponential suppression between the two terms. We will be interested precisely in this regime $\a=\omega/H\gg 1$ in which many background oscillations take place per Hubble time, so we will ignore the exponentially suppressed term of the solution. In the following we call this the \textit{resonance approximation},
\begin{align} \label{resapprox}
    \mathcal{I}_m (k_T) = - \, e^{\pi\a/2} \frac{i^m}{k_T^{m+i\a}}  \Gamma(m+i\alpha) \left[1+ \O(e^{-\pi\a}) \right] \quad \text{(resonance approximation)}\,.
\end{align}

From~\eqref{eq:f} and~\eqref{eq:masterintres} we see that the resonance approximation of the result of the integral~\eqref{eq:ints} can be written as
\beq
    \psi_3 (k_a) = \frac{\text{Poly}_{p+3}(k_a)}{k_T^{p+i\a}} \, ,
\eeq
where $p$ is related to the highest power of $\eta$ appearing in the sum~\eqref{eq:f}. The coefficients of the homogeneous polynomial $\text{Poly}_{p+3}$ (which should not be confused with the polynomials in~\eqref{eq:f}) are complex, and its degree $p+3$ is fixed by the discrete scale invariance condition~\eqref{eq:avgsi}. 

%%%%

 For all tree-level diagrams, $p$ is given by \cite{BBBB}
\begin{align}\label{enri}
    p=1+\sum_A \, (\Delta_A-4)\,,
\end{align}
where $\Delta_A$ is the mass dimension of interaction $A$ and the sum runs over all interactions involved in a given diagram. Notice that this formula also applies to the order of the singularity of partial energy poles, but we will not need this here. For the contact bispectrum that we are studying here, we can easily understand this formula as follows. There is an overall $\eta^{-4}$ factor coming from the integral measure, an $\eta^{n_t+n_i}$ factor coming from the metric contractions of the $n_t$ time and $n_i$ spatial derivatives, and a maximum factor of $\eta^3$ coming from the three propagators~\eqref{eq:Ktimed}. Hence,
\beq
	p-1 = -4+n_t+n_i+3 \qquad\Rightarrow\qquad p = n_t+n_i \, ,
\eeq
So $p$ counts the total number of derivatives of the vertex. This agrees with \eqref{enri} when we notice that there is a single cubic vertex of mass dimension $\Delta=3+n_t+n_i$.\\ %The terms with $1/\eta$ in the polynomial~\eqref{eq:fform} come from the terms in the propagators~\eqref{eq:Ktimed} that contain lower powers of $\eta$.\\

%%%%%%%%%%%%%%%%%%

Before concluding a comment is in order. In previous references \cite{Flauger:2010ja}, the integral in \eqref{Iint} was in practice approximated using the stationary phase approximation\footnote{The method of stationary phase says that as $x\to\infty$
\beq
	\int_a^b g(t) \, e^{ix \varphi (t)} \, \rd t  = \left( \frac{2\pi}{x|\varphi''(c)|} \right)^{1/2} \cdot g(c) \cdot \exp{\left[ ix\varphi(c) + i \frac{\pi}{4} \rm{sgn}\left( \varphi''(c) \right) \right]} + O (x^{-3/2}) \, ,
\eeq
where the phase $\varphi (t)$ has a stationary point $\varphi'(c)=0$ at $a<c<b$. Note that if $g(c)\sim O(x^n)$ with $n\neq 0$, the error becomes $O(x^{n-3/2})$. \label{ft:spa}}
(even though the exact gamma function expression was also derived). Here we do \textit{not} invoke such approximation and our resonance approximation includes all power law suppressed terms in $1/\alpha$, which would be missed by the leading order stationary phase result. As we will see shortly, these subleading terms beyond the stationary phase approximation are actually fixed by locality in the form of the manifestly local test. \\

In summary, all wavefunction coefficients coming from tree-level bulk interactions in dS as in \eqref{eq:ints} take the form of our ansatz \eqref{simply} up to exponentially suppressed terms in $e^{-\pi \alpha}$ (resonance approximation). This ansatz provides a rich class of minimal solutions to the discrete scale invariance constraint in \eqref{eq:avgsi}.

%%%%%%%%%%%%%%%%%%%%
\subsection{Bose symmetry} \label{sec:Bose}

We are computing wavefunction coefficients of a bosonic field $\phi$, so the result must be symmetric under any permutation of the field labels. Since the oscillatory part of \eqref{eq:factorized} is symmetric, this implies that $\hat{\psi}_3$ must also be so, %and since the different terms in~\eqref{eq:psihat2} come with different powers of $\a/k_T$, each of the polynomials $\text{Poly}^{(m)} (k_a)$ must be separately symmetric. This 
which in turn implies that $\text{Poly}_{p+3} (k_a)$ can be written in a unique way in terms of the three elementary symmetric polynomials
\begin{align}
 k_T&\equiv k_1+k_2+k_3\,, & e_2&\equiv k_1k_2+k_2k_3+k_3k_1\,, & e_3&\equiv k_1k_2k_3 \,.
\end{align}
Hence
\beq \label{eq:psihatsym}
	% \hat{\psi}_3 (k_a) = \left( \frac{i\a}{k_T} \right)^p \text{Poly}^{(p)}_{p+3} (k_T,e_2,e_3) + \left( \frac{i\a}{k_T} \right)^{p-1} \text{Poly}^{(p-1)}_{p+2} (k_T,e_2,e_3) \, .
    % \hat{\psi}_3 (k_a) = \left( \frac{i\a}{k_T} \right)^p \text{Poly}_{p+3} (k_T,e_2,e_3) \, ,
    \hat{\psi}_3 (k_a) = \frac{\text{Poly}_{p+3} (k_T,e_2,e_3)}{k_T^p} \, ,
\eeq
and the whole wavefunction coefficient is 
\beq \label{eq:psisym}
	% \psi_3 (k_a) = \left[  \left( \frac{i\a}{k_T} \right)^p \text{Poly}^{(p)}_{p+3} (k_T,e_2,e_3) + \left( \frac{i\a}{k_T} \right)^{p-1} \text{Poly}^{(p-1)}_{p+2} (k_T,e_2,e_3) \right] \cdot e^{-i\a\log{(k_T/k_*)}} + O(\a^{p-2}) \, .
    % \psi_3 (k_a) = \left( \frac{i\a}{k_T} \right)^p \text{Poly}_{p+3} (k_T,e_2,e_3) \cdot e^{-i\a\log{(k_T/k_*)}} \, .
    \psi_3 (k_a) = \frac{\text{Poly}_{p+3} (k_T,e_2,e_3)}{k_T^{p+i\a}} \, .
\eeq

%%%%%%%%%%%%%%%%%%%%
\subsection{Cosmological optical theorem} \label{sec:COT}

The Cosmological Optical Theorem (COT) \cite{COT} still applies in the presence of time dependent couplings such as the ones appearing in resonant models, as long as they are real (and hence the theory is unitary) and the state of the universe approaches the Bunch-Davies state in the infinite past. This implies that the exact wavefunction coefficients that we are interested in must satisfy the COT for contact diagrams, namely
\beq \label{eq:COT}
\text{Cosmo optical theorem: }\quad \psi_3^{\text{exact}} (k_a) + \left[ \psi_3^{\text{exact}} (-k_a^*) \right]^* = 0 \, , \qquad k_a\in \mathbb{C}^{n-} \, .
\eeq
Indeed the exact propagators appearing in the presence of an oscillatory background do obey hermitian analyticity \cite{Goodhew:2021oqg} and this ensures that the proof of the COT proceeds in the same way as described in \cite{sCOTt}. Moreover, notice that this relation is valid at all times and hence also applies in the case of IR divergences when we need to introduce an IR regulator, see Section \ref{ss:IR}.\\

However there is a practical problem: in our approach we are not computing the \textit{exact} wavefunction coefficient $\psi_3^{\text{exact}}$. 
Rather we aim to capture its resonance approximation in \eqref{resapprox}, by neglecting terms that are exponentially suppressed in $\alpha$.  The issue is that the analytic continuation to negative energies, $k\to-k$, present in the COT, turns an exponentially suppressed term such as $e^{-\pi \alpha}e^{i\alpha \log{k_T}}$ into an unsuppressed term. The resulting term ends up canceling out with the leading term to satisfy \eqref{eq:COT}. In the resonance approximation we neglect the exponentially suppressed contribution and so this cancellation cannot be checked or used\footnote{Incidentally, we notice that also the stationary phase approximation, which we don't use, is incompatible with the  COT. Indeed only one of the two saddle points is within the domain of integration, but the saddles are interchanged by the analytic continuation $k_a\to -k_a^*$ appearing in the COT. Indeed $\eta_{\rm res} = \pm \a/k_T $ turns into $\eta_{\rm res} = \mp \a/k_T^* $, which for $k_a\in\mathbb{R}^+$ means that $\eta_{\rm res}$ is now outside (inside) of the domain of integration for the integral with the minus (plus) sign.}. Another way to see the problem is to notice that the resonance approximation trades the real coupling constant $cos(\omega t)$ for the complex coupling constant $e^{i \omega t}$, hence invalidating the assumptions of unitarity. Throughout this paper we denote the resonance approximation simply by 
\begin{align}
 \psi_3\equiv\psi_3^{\text{resonance approx.}}\,.
\end{align}
Having identified this problem, there are two possible solutions. One is to modify the ansatz including the exponentially suppressed terms that are being neglected, in such a way that the solution satisfies~\eqref{eq:COT}. The other is to find a modified version of the COT that is satisfied by the resonance approximation. The latter turns out to be the most straightforward approach and the one we choose to follow. To find a modified COT notice that our effective coupling $e^{-i\alpha H t}/2$ remains unchanged if we take its complex conjugate \textit{and} send $\alpha\to -\alpha$. In other words, it behaves just as a real coupling would under complex conjugation. This in turn ensures that the derivation of the COT of \cite{sCOTt} goes through\footnote{More in detail, the propagator identities are unchanged and we can still trade the complex conjugation of the propagators for an analytic continuation of the energies \textit{and} a flipping of the sign of $\alpha$ to get a relation for the wavefunction coefficients.} and applies to the resonance approximation. Hence, for contact diagrams we impose the following
\beq \label{eq:mCOT}
	\text{Modified cosmo optical theorem: } \qquad \psi_3 (k_a,\alpha) + \left[ \psi_3 (-k_a^*,-\alpha) \right]^* = 0 \, , \qquad k_a\in \mathbb{C}^{n-} \, ,
\eeq
where $\a\in\mathbb{R}^+ $ and we have made the $\alpha$-dependence explicit. It would be interesting to check this modified relation to all orders in perturbation theory, but we will leave that for the future.\\

To apply the modified COT to our solution~\eqref{eq:psisym} we need to take into account the overall factor of $e^{\pi\a/2}$ that results from the time integrals, cf. Equation~\eqref{eq:masterintres}, and that we are ignoring. We have absorbed this exponential factor into an overall normalization that we are not fixing, but it is essential for the COT to work and so we temporarily restore it here\footnote{Notice that gamma functions $\Gamma(m+i\alpha)$ remain invariant so we can omit them.}:
\beq
    \psi_3 (k_a,\a) = e^{\pi\a/2} \, \frac{\text{Poly}_{p+3} (k_T,e_2,e_3)}{k_T^{p+i\a}} \, .
\eeq
For real momenta the conjugate wavefunction is then
\beq
    \left[ \psi_3 (-k_a^*,-\a) \right]^* = - \, e^{\pi\a/2} \frac{\left[ \text{Poly}_{p+3} (k_T,e_2,e_3,-\a) \right]^*}{k_T^{p+i\a}} \, ,
\eeq
and plugging this into the modified COT~\eqref{eq:mCOT} we get
\beq \label{eq:COTpoly}
	\text{Poly}_{p+3} (k_T,e_2,e_3,\a) = \left[ \text{Poly}_{p+3} (k_T,e_2,e_3,-\a) \right]^* \, .
\eeq
At a given order $p$ in derivatives, we can write the polynomial explicitly as follows,
\beq \label{eq:poly}
    \text{Poly}_{p+3} (k_T,e_2,e_3,\a) = \sum_{s=0}^{\lfloor\frac{p+3}{3}\rfloor} \sum_{r=0}^{\lfloor\frac{p+3-3s}{2}\rfloor} C_{r,s} (\a) \, k_T^{p+3-2r-3s} \, e_2^r \, e_3^s \, ,
\eeq
where $\lfloor\ldots\rfloor$ is the floor function and the coefficients $C_{r,s}$ can depend on any parameter of the theory but here only the $\a$-dependence is relevant for our purposes. Since~\eqref{eq:COTpoly} must be satisfied for any real kinematic configuration, we can match both sides of the equation individually for every monomial in $\{k_T,e_2,e_3\}$ that appears in them. What we get is that coefficients %with an even power of $\a$ must be purely real, while those with an odd power must be purely imaginary. 
$C_{r,s}$ must satisfy
\beq \label{eq:COTcoefs}
    C_{r,s} (\a) = C_{r,s}^* (-\a) \, ,
\eeq
this is, any appearance of $\a$ in them must come with a factor of $i$. %, so $\text{Poly}_{p+3}$ is a polynomial on $\{k_T,e_2,e_3,i\a\}$ with real coefficients. 
%The result is then that \textit{all the coefficients of $\text{Poly}^{(p)}$ and $\text{Poly}^{(p-1)}$ must be real},
% \begin{align}
%     \text{Poly}^{(p)},\text{Poly}^{(p-1)}\in \mathbb{R}\,.
% \end{align}
Remember that we are ignoring an overall complex factor in $\psi_3$, whose phase we have absorbed in the definition of $k_*$, so what unitarity really dictates is that there is a \textit{relative} phase of $\pi/2$ between successive powers of $\a$ in the wavefunction coefficient. In the following, we will not write the $\a$-dependence of the coefficients $C_{r,s}$ explicitly to avoid clutter.

%%%%%%%%%%%%%%%%%%%%
\subsection{Manifestly local test} \label{sec:MLT}

To capture local models we are interested in wavefunction coefficients for massless scalars that come from manifestly local interactions, namely interactions of the product of fields and their derivatives at the same spacetime point. In particular, the derivation in Section 3.2 of~\cite{MLT} applies in this case and we conclude that our wavefunction coefficients must satisfy the so-called Manifestly Local Test (MLT)
\beq \label{eq:MLT}
	\left. \frac{\del}{\del k_1} \, \psi_3 (k_a) \right|_{k_1\to 0} = 0 \, ,
\eeq
where Bose symmetry allows us to pick the energy $k_1$ without loss of generality. It will be useful to notice that the different orders in $\a$ of the ansatz~\eqref{eq:psisym} do not need to separately satisfy the MLT, since the derivative $\del/\del k_1$ acting on the oscillatory factor mixes them. The MLT thus relates different orders in $\a$ of the solution. Notice that, just like the COT, the MLT applies to all finite times as well and so can also be used in the presence of IR divergences. \\

If we demand our solution~\eqref{eq:psisym} to satisfy the MLT~\eqref{eq:MLT}, we get
\begin{equation} \label{eq:MLTexp}
\begin{aligned}
    0 & = \left. \frac{\del}{\del k_1} \, \psi_3 (k_a) \right|_{k_1\to 0} = \\
    & = \left. \left[ \frac{\del}{\del k_1} \, \text{Poly}_{p+3} (k_T,e_2,e_3) -\frac{i\a + p}{k_T} \, \text{Poly}_{p+3} (k_T,e_2,e_3) \right] \frac{1}{k_T^{p+i\a}} \right|_{k_1\to 0} \, .
\end{aligned}
\end{equation}
Now we can plug the polynomial expansion~\eqref{eq:poly} into this expression and solve for the coefficients $C_{r,s}$. Since the MLT must be satisfied for general kinematics, we have to demand every monomial in $\{k_2,k_3\}$ of~\eqref{eq:MLTexp} to vanish. At order $p$ in derivatives, this results in the following set of relations between coefficients:
\beq \label{eq:coefsMLT}
    \left\{ \begin{aligned}
    (i\a -3) \, C_{0,0} & = C_{1,0} \, , \\
    (i\a -3 +2r) \, C_{r,0} & = (r+1) \, C_{r+1,0} + C_{r-1,1} \qquad \text{for $1\leq r \leq \lfloor \frac{p+1}{2} \rfloor$} \, , \\
    (i\a+p-1) \, C_{\frac{p+2}{2},0} & = C_{\frac{p}{2},1} \qquad \text{if $p$ is even} \, , \\
    C_{\frac{p+3}{2},0} & = 0 \qquad \text{if $p$ is odd} \, .
    \end{aligned} \right.
\eeq
In the next section we will use this to write the general solution.

%%%%%%%%%%%%%%%%%%%%%%%%%%%%%%%%%%%%%%%%%%%%%%%%%%%%%%%%%%%%%%%%%%
\section{Bootstrapping resonant bispectra}\label{sec:bootstrap}

In Section~\ref{sec:rules} we have applied the bootstrap rules and we have obtained the most general three-point wavefunction coefficient under our assumptions, which is given by the wavefunction~\eqref{eq:psisym} with a polynomial ~\eqref{eq:poly} whose coefficients satisfy~\eqref{eq:COTcoefs} and~\eqref{eq:coefsMLT}. Nevertheless, this expression contains redundancies, and in the next subsection we will eliminate them in order to present the result in a clearer way. After that, we will perform two consistency checks on our expressions---we will study the flat space limit and count the number of structures. Then we will explicitly write down the lowest-$p$ results for the shapes of resonant non-Gaussianity. Finally, we will comment on parity-odd interactions and IR-divergent terms.

%%%%%%%%%%%%%%%%%%%%
\subsection{Final result} \label{sec:final}

Our result  $\psi_3 (k_a)$, for a given order $p$ in derivatives, is given by~\eqref{eq:psisym} with a polynomial~\eqref{eq:poly} whose coefficients satisfy~\eqref{eq:COTcoefs} and~\eqref{eq:coefsMLT}. Notice that such a formula contains the new kinematic structures that appear at order $p$ in derivatives (those with $k_T^0$ terms in $\text{Poly}_{p+3}$) plus all the structures corresponding to lower $p$ (those with an overall nonzero power of $k_T$ in $\text{Poly}_{p+3}$). It is desirable to have an expression that just provides the new wavefunction coefficients $\psi_3^{(p)} (k_a)$ at each order $p$ in derivatives. Before imposing the MLT constraints~\eqref{eq:coefsMLT}, the new structures that appear in $\text{Poly}_{p+3}$ at each order $p$ are
\beq \label{eq:newstrs}
    \sum_{s=0}^{\lfloor\frac{p+3}{3}\rfloor} C_{\frac{p+3-3s}{2},s} \cdot \delta_{\lfloor\frac{p+3-3s}{2}\rfloor,\frac{p+3-3s}{2}} \cdot e_2^{\frac{p+3-3s}{2}} \, e_3^s \, ,
\eeq
where the Kronecker delta ensures that the power of $e_2$ is integer. Let us now see how the MLT~\eqref{eq:coefsMLT} ties some of the coefficients of these new structures to other coefficients in $\text{Poly}_{p+3}$. Notice first that there is an $s=0$ term in~\eqref{eq:newstrs} only when $p$ is odd, but in that case \eqref{eq:coefsMLT} requires the corresponding coefficient to vanish, $C_{\frac{p+3}{2},0} = 0$. We can then just start the sum in~\eqref{eq:newstrs} at $s=1$:
\beq
    \sum_{s=1}^{\lfloor\frac{p+3}{3}\rfloor} C_{\frac{p+3-3s}{2},s} \cdot \delta_{\lfloor\frac{p+3-3s}{2}\rfloor,\frac{p+3-3s}{2}} \cdot e_2^{\frac{p+3-3s}{2}} \, e_3^s \, .
\eeq
Among these monomials, those with $s>1$ are unconstrained by the MLT~\eqref{eq:coefsMLT}, so we will write them in the final solution with a free coefficient. On the other hand, the structure with $s=1$---which is present only when $p$ is even---is tied by the MLT~\eqref{eq:coefsMLT} to other structures with different powers of $k_T$ as follows. For $p=0$ we have
\begin{equation}
    \left. \begin{aligned}
        C_{1,0} & = (i\a-3) \, C_{0,0} \\
        C_{0,1} & = (i\a-1)(i\a-3) \, C_{0,0} 
    \end{aligned} \right\} \, ,
\end{equation}
while for even $p>0$ we have
\begin{equation}
    \left. \begin{aligned}
        C_{\frac{p}{2},1} & = (i\a+p-1) \, C_{\frac{p+2}{2},0} \\
        C_{\frac{p-2}{2},1} & = (i\a+p-3) \, C_{\frac{p}{2},0} - \frac{p+2}{2} \, C_{\frac{p+2}{2},0} \\
    \end{aligned} \right\} \, .
\end{equation}
With this information, we are ready to state the main result of this paper. The independent shapes that contribute to the resonant, tree-level, three-point wavefunction coefficient at order $p$ in derivatives are given by 
\begin{tcolorbox}[colframe=white,arc=0pt,colback=greyish2]
\begin{equation} \label{eq:finalres}
    \psi_3^{(p)} = \frac{1}{k_T^{p+i\a}} \times \left\{ \begin{aligned}
        & C_{0,0} \left[ (i\a-1)(i\a-3) e_3 + (i\a-3) e_2k_T + k_T^3 \right] \qquad\quad \text{for $p=0$} \, , \\[10pt]
        & \sum_{s=2}^{\lfloor\frac{p+3}{3}\rfloor} C_{\frac{p+3-3s}{2},s} \cdot \delta_{\lfloor\frac{p+3-3s}{2}\rfloor,\frac{p+3-3s}{2}} \cdot e_2^{\frac{p+3-3s}{2}} \, e_3^s  \qquad\quad \text{for odd $p$} \, , \\[10pt]
        & C_{\frac{p+2}{2},0} \cdot e_2^{\frac{p-2}{2}} \left[ (i\a+p-1) \, e_2 e_3 - \frac{p+2}{2} k_T^2 e_3 + k_T e_2^2 \right] \\[-4pt]
        & + \sum_{s=3}^{\lfloor\frac{p+3}{3}\rfloor} C_{\frac{p+3-3s}{2},s} \cdot \delta_{\lfloor\frac{p+3-3s}{2}\rfloor,\frac{p+3-3s}{2}} \cdot e_2^{\frac{p+3-3s}{2}} \, e_3^s  \qquad\quad \text{for even $p>0$} \, .
    \end{aligned} \right.
\end{equation}
\end{tcolorbox}
\noindent In this context, the most general three-point wavefunction coefficient will then be given by a linear combination of these shapes, with coefficients $C_{r,s}$ that must satisfy the COT constraint derived in Subsection~\ref{sec:COT}, namely that they must be invariant under the combined action of complex conjugation and $\a\to -\a$.

%%%%%%%%%%%%%%%%%%%%
\subsection{Flat space limit} 

The ansatz for $\psi_3(k_a)$ generically has a total energy pole of order $p$ equal to the number of derivatives of the operator at hand. This is the same total energy pole encountered in usual scale-invariant wavefunction coefficients~\cite{Maldacena:2011nz,Raju:2012zr}, and its residue is also given by the high-energy limit of a corresponding scattering amplitude. The argument is similar to that of the usual case---the structure of the bulk integral~\eqref{eq:ints} implies that its result when $k_T\to 0$ is approximated by the integrand times $\eta$ evaluated at $\eta%=\eta_{\rm res}
\to -\infty$, a limit in which we recover the amplitude. Ignoring the phase from the ${\rm exp} (-i\a\log{(k_T/k_*)})$ oscillatory factor, we have then
\beq \label{eq:flatspace}
	\lim_{k_T\to 0} \hat{\psi}_3^{(p)} \sim \frac{e_3 \, A_3^{(p)}(k_a)}{k_T^p} \, .
\eeq
The three-point amplitude $A_3^{(p)}(k_a)$ comes from a vertex with $p$ derivatives and hence has dimension $p$. %It will be an input of our bootstrap procedure. 
Since we are considering boost-breaking interactions, we will be interested in the most generic scattering amplitude coming from $p$-derivative vertices that break Lorentz boosts, which is given by the most general polynomial of $\{e_2,e_3\}$ of degree $p$~\cite{PSS},
\beq \label{eq:amp}
	A_3^{(p)}(k_a) = \text{Poly}_{p} (e_2,e_3) \, .
\eeq
If we look at the $k_T\to 0$ limit of the general result~\eqref{eq:finalres}, we indeed find~\eqref{eq:flatspace} with the amplitudes~\eqref{eq:amp}. This is a nice consistency check of our result, % and can be recast in more colorful language. 
but there is more to it. Notice that our results~\eqref{eq:finalres} satisfy
\beq \label{eq:bigalphalim}
    \psi_3^{(p)} = \frac{e_3 \, A_3^{(p)}(k_a)}{k_T^{p+i\a}} + \mathcal{O}\left( \frac{1}{\a} \right) \, .
\eeq
When the couplings are constants, non-Gaussian correlators are generated around horizon exit (except in the few cases with IR divergent interactions). Around horizon exit, the effect of the expansion are not negligible and the wavefunction coefficients or the corresponding correlators contain both a term corresponding to the flat-space amplitude \textit{and} other curved spacetime contributions. While these two contributions can be separated using complex momenta and the \textit{mathematical} limit $k_T \to 0$, for general \textit{physical} configurations they are of the same order and generate a similar amount of non-Gaussianity. The resonant case stands in stark contrast to this expectation, and indeed can be thought of a way to zoom into the flat space limit by taking the large limit of the \textit{physical} parameter $\alpha $, namely the frequency of oscillations in units of the Hubble parameter. In the limit $\alpha \gg 1$, the non-Gaussianity is produced on scales parametrically shorter that Hubble and so is fully captured by flat space physics, up to higher orders in $O(1/\alpha)$. In other words, to leading order in $\alpha$, resonant non-Gaussianities are just Minkowski amplitudes dressed with the oscillatory factor $e^{-i\alpha \log k_T}$!\\

%%%%%%%%%%%%%%%%%%%%
\paragraph{Counting structures}

Here we count the number of possible structures that result from the bootstrap process up to a given order $p$, and prove that the total number is equal to that of \textit{independent} cubic vertices with at most $p$ derivatives. The latter can be found by counting the structures in the amplitude~\eqref{eq:amp}, and is equal to the number of non-negative integer solutions $\{ r,s \}$ to $2r+3s\leq p$. This number is~\cite{Mahmoudvand}
\beq
	N_{\rm amp} (p) = \sum_{q=0}^{\lfloor\frac{p}{2}\rfloor} \left( 1+\lfloor\frac{p-2q}{3}\rfloor \right) \, .
\eeq

The result~\eqref{eq:finalres} introduces, at every order in $p$, a number of new structures parameterised by the real constants $C_{r,s}$. These constants come precisely from the amplitude~\eqref{eq:amp}, as shown in~\eqref{eq:bigalphalim}, so their number up to $p$ derivatives will be exactly $N_{\rm amp} (p)$. 
We conclude then that the number of independent structures in $\psi_3$ up to order $p$ is $N_{\rm amp} (p)$, and hence it coincides with the number of cubic vertices with at most $p$ derivatives.

Notice the difference with the standard bootstrap case, in which the number of structures was equal to the number of scattering amplitudes plus one~\cite{MLT}. The extra structure corresponded to the local non-gaussianity associated to the only manifestly local field redefinition, $\f\to\f+\f^2$, which has the form $\psi_3^{\rm local} = \psi_2 (k_1) + \psi_2 (k_2) + \psi_2 (k_3) = k_T^3-3k_Te_2+3e_3$. In the resonant case, however, we have\footnote{This two-point wavefunction coefficient can be obtained from a bulk computation without much difficulty, but Appendix~\ref{app:anstr} presents a way of bootstrapping its functional form purely from the boundary.}
\beq \label{eq:respsi2}
	\psi_2 (k) \sim k^3 \Bigg( 1 + i \, 3b_*  \sqrt{\frac{2\pi}{\a}} e^{ i\a \log{( k/k_* )} } \Bigg) \, ,
\eeq
with $b_*<1$ a parameter that controls the monotonicity of the potential. The corresponding sum $\psi_2 (k_1) + \psi_2 (k_2) + \psi_2 (k_3)$ cannot be captured by our ansatz~\eqref{eq:factorized} for $\psi_3$ and consequently we do not find in our analysis a three-point wavefunction coefficient coming from the above field redefinition.

%%%%%%%%%%%%%%%%%%%%
\subsection{Comparison with previous results} \label{sec:compa}

The literature on resonant non-gaussianities has traditionally presented the bispectrum results as an expansion in $\a\gg 1$, often computing just the leading order but sometimes providing the NLO and even the NNLO corrections too. The overall phase of the oscillations---which we called $k_*$ in our approach---is an undetermined parameter because it depends on the background solution and on the expansion history after inflation. Here we recall that a shift of said phase is degenerate with a change in the functional dependence of the subleading terms in the $\a$ expansion. To see this, consider a resonant bispectrum of the form
\beq
    B(k_a) = f(k_a) \cos\Big( \a\log (k_T/k_*) \Big) + \frac{1}{\a} \cdot g(k_a) \sin\Big( \a\log (k_T/k_*) \Big) \, ,
\eeq
where $k_a$ collectively denotes $\{ k_1,k_2,k_3 \}$ and $f(k_a)$ and $g(k_a)$ are rational functions independent of $\a$. The overall phase is given by $k_*$ and is undetermined, so we are free to shift it. If we shift it by an $\a$-dependent amount, like for example $\Delta\varphi = \arctan (1/\a)$, we have
\begin{equation}
    \begin{aligned}
        B(k_a) & = f(k_a) \cos\Big( \a\log (k_T/k_*) +\Delta\varphi \Big) + \frac{1}{\a} \cdot g(k_a) \sin\Big( \a\log (k_T/k_*) +\Delta\varphi \Big) \\
        & = \frac{\a}{\sqrt{1+\a^2}} \left[ f(k_a) \cos\Big( \a\log (k_T/k_*) \Big) + \frac{1}{\a} \Big( g(k_a) - f(k_a) \Big) \sin\Big( \a\log (k_T/k_*) \Big) \right. \\
        & \qquad\qquad\qquad \left. +\frac{1}{\a^2} \cdot g(k_a) \cos\Big( \a\log (k_T/k_*) \Big) \right] \, .
    \end{aligned}
\end{equation}
Notice that the kinematic dependence of the NLO term of the expansion has now changed, and a new term has appeared at NNLO. This fact, which was already pointed out in~\cite{Leblond:2010yq} to explain an apparent discrepancy with the result of~\cite{Flauger:2010ja}, is general: the form of the subleading terms in the $\a$-expansion can be changed in a specific way by an $\a$-dependence change of the overall phase.

Fortunately, this degeneracy is inconsequential for phenomenology since the overall phase of the oscillations is marginalised when these signals are looked for in the data. However, it will be important when comparing our bootstrap results to the previous literature. Recall that our approach does not determine the overall size and phase of the wavefunction coefficients, so comparison must always be done up to overall factors containing powers of $i\a$ (as per the COT constraint of Section~\ref{sec:COT}) or, equivalently, up to a phase shift (for instance, the phase shift $\Delta\varphi = \arctan (1/\a)$ in the above example can be understood as coming from an overall factor of $(1-1/i\a)\,\psi_3$ in the wavefunction coefficient). %In this sense, the bootstrap solutions are given in the basis of structures that satisfy the MLT, cf.~\eqref{eq:finalres}, which is different from

Finally, let's mention that resonant non-Gaussianity from interactions with a small number of derivatives, as indeed the case in the original example \cite{Flauger:2010ja}, induce IR divergences in the wavefunction coefficient, which had not been previously discussed in the literature. We make up for this omission in Section \ref{ss:IR}.

%%%%%%%%%%%%%%%%%%%%
\subsection[Lowest-$p$ shapes]{Lowest-$\boldsymbol{p}$ shapes} \label{sec:lowestp}

We will now explicitly write all the possible resonant bispectra at increasing order $p$ of the total energy pole, up to $p=3$. Recall that the relation between the bispectrum and the wavefunction is given by
\beq
    B(k_1,k_2,k_3) = \frac{{\rm Re} \, \psi_3 (k_1,k_2,k_3)}{4 \prod_a {\rm Re}\, \psi_2 (k_a)} \, .
\eeq
In some cases, we will single out certain linear combinations that correspond to wavefunction coefficients coming from specific vertices in the Effective Field Theory of Inflation (EFToI). In the context of resonant inflation the continuous shift symmetry of the Goldstone boson $\pi$ becomes a discrete shift symmetry, and the couplings acquire an oscillatory time dependence~\cite{Behbahani:2011it,Behbahani:2012be}. Schematically, the wavefunction coefficients that we will find correspond to interaction terms in the Lagrangian of the form
\beq
    \mathcal{L} \quad \supset \int dt d^3x \,a^3 \, \cos{(\omega t)} \left[ \lambda_1 \pi^3 + \lambda_2 \pi^2 \dot{\pi} + \lambda_3 \pi(\del_i\pi)^2 +\lambda_4 \pi \dot \pi^2 +  \ldots \right]\,,
\eeq
where the phase and overall size of the couplings are immaterial since they are not captured by the bootstrap procedure. Here we focus on the IR finite part of the result, and we discuss IR divergences in Section \ref{ss:IR}

%%%%%%%%%
\subsubsection*{$\boldsymbol{p=0}$}

\begin{figure}[h!]
\centering
\begin{subfigure}{0.42\textwidth}
    \includegraphics[width=\textwidth]{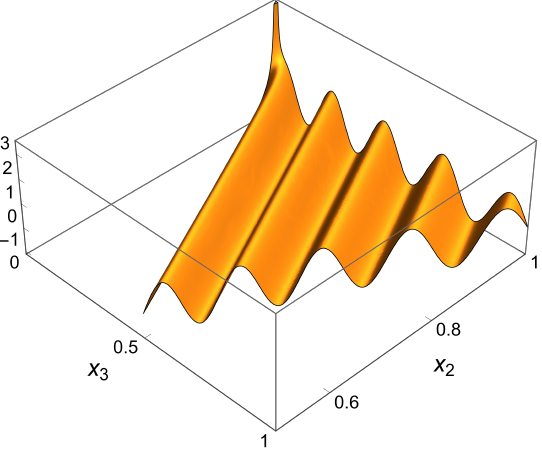}
    \caption{Shape of the $p=0$ bispectrum $B^{(0)}$~\eqref{eq:Bp0}.}
    \label{fig:plotp0}
\end{subfigure}
\hfill
\begin{subfigure}{0.42\textwidth}
    \includegraphics[width=\textwidth]{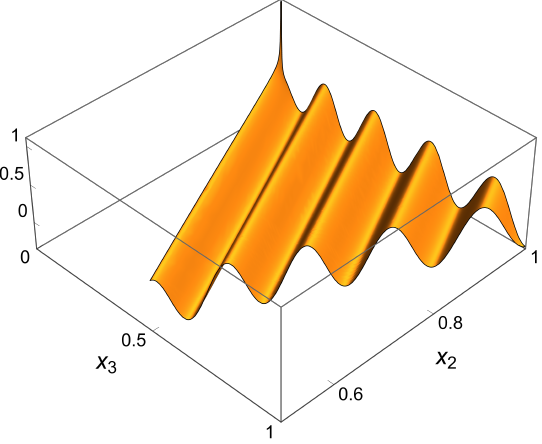}
    \caption{Shape of the $p=2$ bispectrum $B^{(2)}$~\eqref{eq:Bp2}.}
    \label{fig:plotp2}
\end{subfigure}
\hfill
\begin{subfigure}{0.42\textwidth}
    \includegraphics[width=\textwidth]{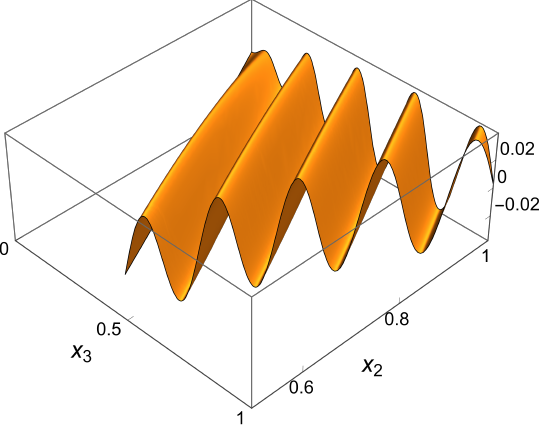}
    \caption{Shape of the $p=3$ bispectrum $B^{(3)}$~\eqref{eq:Bp3}.}
    \label{fig:plotp3}
\end{subfigure}
\caption{Shape $S(k_1,k_2,k_3)\equiv (k_1k_2k_3)^2 B(k_1,k_2,k_3)$ of three resonant bispectra plotted as a function of $x_a\equiv k_a/k_1$ for $k_*=k_1$ and $\a=70$. Notice that due to the absence of scale invariance these plots change under a rescaling $k_a\to \lambda k_a$ or, equivalently, under a change of phase $k_*$.}
\label{fig:plots}
\end{figure}

The result~\eqref{eq:finalres} for $p=0$ is simply
\beq \label{eq:resp0}
	\psi_3^{(0)} = C_{0,0} \left( \frac{k_*}{k_T} \right)^{i\a} \left[ (i\a-1)(i\a-3) e_3 + (i\a-3) e_2k_T + k_T^3 \right] \, ,
\eeq
where we have re-introduced the overall phase $k_*$. The bispectrum is then, up to an overall real factor,
\begin{equation} \label{eq:Bp0}
\begin{aligned}
	B^{(0)} (k_1,k_2,k_3) \propto \frac{1}{(k_1k_2k_3)^2} & \left[ \cos\Big( \a\log (k_T/k_*) \Big) + \frac{1}{\a} \left( 1 - \sum_{i\neq j} \frac{k_i}{k_j} \right) \sin\Big( \a\log (k_T/k_*) \Big) \right. \\
    & \left. -\frac{1}{\a^2} \cdot \frac{k_1^3+k_2^3+k_3^3}{k_1k_2k_3} \cos\Big( \a\log (k_T/k_*) \Big) \vphantom{\left( 1 - \sum_{i\neq j} \frac{k_i}{k_j} \right)} \right] \, .
\end{aligned}
\end{equation}
Figure~\ref{fig:plotp0} shows a slice of the corresponding shape $S(k_1,k_2,k_3) \equiv (k_1k_2k_3)^2 B(k_1,k_2,k_3)$. A bulk computation confirms that~\eqref{eq:Bp0} corresponds to the bispectrum arising from a vertex $\pi^3$ (up to a suitable choice of phase $k_*$ as explained in Subsection~\ref{sec:compa}):
\beq \label{eq:Bpicube}
    B_{\pi^3} (k_1,k_2,k_3) = B^{(0)} (k_1,k_2,k_3) \, .
\eeq
This bispectrum was computed in~\cite{Leblond:2010yq,Behbahani:2011it} by evaluating the time integrals at the stationary phase point, but the subleading corrections in $\a\gg 1$ were not correctly taken into account, thus getting an incorrect result at $\mathcal{O}(1/\a)$. 
The symmetric squeezed limit of this bispectrum is\footnote{By symmetric squeezed limit of a bispectrum $B(k_1,k_2,k_3)$ we mean
\beq
    \lim_{|\vk_l|\to 0} B \Big( |\vk_l|,\left| \vk_s-\vk_l/2 \right| ,\left| -\vk_s-\vk_l/2 \right| \Big) \, ,
\eeq
where $\vk_l$ and $\vk_s$ are the long and short modes respectively. The result is generically expressed in terms of $k_l\equiv |\vk_l|$, $k_s\equiv |\vk_s|$ and $\cos\theta \equiv \hat{\vk}_l\cdot\hat{\vk}_s$.}
\beq \label{eq:slp0}
    \lim_{|\vk_l|\to 0} B^{(0)} = - \frac{2}{\a \, k_s^3 \, k_l^3} \left[ \sin ( \a\log (2k_s/k_*) ) + \frac{1}{\a} \cos ( \a\log (2k_s/k_*) ) \right] + \mathcal{O} \left( \frac{1}{k_l} \right) \, .
\eeq
It was argued in~\cite{Creminelli:2011rh,Assassi:2015jqa} that the term that diverges as $\mathcal{O} (1/k_l^2)$ must be absent in the symmetric squeezed limit. This was checked explicitly for a resonant model in~\cite{Creminelli:2011rh}, although the bispectrum was obtained via the stationary phase approximation and the cancellation only worked at leading and next-to-leading orders in $1/\a$. For our bootstrap solution, however, Equation~\eqref{eq:slp0} shows that the absence of a $\mathcal{O} (1/k_l^2)$ term is exact.

%%%%%%%%%
\subsubsection*{$\boldsymbol{p=1}$}

We can see from~\eqref{eq:finalres} that there is no new wavefunction coefficient for $p=1$. In the EFToI there is one vertex with one derivative, $\pi^2\dot{\pi}$, and from this point of view the corresponding bispectrum can only be given by the structure that we already had at $p=0$, i.e. 
\beq \label{eq:Bpisqpidot}
    B_{\pi^2\dot{\pi}} (k_1,k_2,k_3) = B^{(0)} (k_1,k_2,k_3) \, .
\eeq
A bulk computation confirms that this is indeed the case, and the bispectra coming from $\pi^3$ and $\pi^2\dot{\pi}$ are the same up to an $\a$-dependent phase shift. The bispectrum~\eqref{eq:Bpisqpidot} coincides with the result in~\cite{Flauger:2010ja} at leading order in $1/\a$ but differs starting at next-to-leading order. The reason is again that the bispectrum there was computed via the stationary phase approximation, which gives the correct result only to leading order in $1/\alpha$. The difference however can be re-absorbed into an $\a$-dependent phase and so is phenomenologically inconsequential.

%%%%%%%%%
\subsubsection*{$\boldsymbol{p=2}$}

In this case we have
\beq \label{eq:resp2}
	\psi_3^{(2)}   = C_{2,0} \left( \frac{k_*}{k_T} \right)^{2+i\a} \left[ (i\a+1) \, e_2 e_3 - 2 k_T^2 e_3 + k_T e_2^2 \right] \, ,
\eeq
with bispectrum
\beq \label{eq:Bp2}
	B^{(2)} (k_1,k_2,k_3) \propto \frac{e_2}{e_3^2k_T^2} \sin\Big( \a\log (k_T/k_*) \Big) + \frac{1}{\a} \cdot \frac{1}{e_3k_T^2} \left( \sum_j \frac{k_T}{k_j^2} + \sum_j \frac{1}{k_j} \right) \cos\Big( \a\log (k_T/k_*) \Big) \, .
\eeq
The symmetric squeezed limit is
\beq \label{eq:slp2}
    \lim_{|\vk_l|\to 0} B^{(2)} = \frac{\cos ( \a\log (2k_s/k_*) )}{2 \, \a \, k_s^3 \, k_l^3} + \mathcal{O} \left( \frac{1}{k_l} \right) \, ,
\eeq
where we can see again the absence of the $\mathcal{O}(1/k_l^2)$ term.
The EFToI vertex $\pi\dot{\pi}^2$ gives rise to exactly the wavefunction coefficient~\eqref{eq:resp2}, so its bispectrum is
\beq \label{eq:bisppidotpisq}
	B_{\pi\dot{\pi}^2}  (k_1,k_2,k_3) = B^{(2)} (k_1,k_2,k_3) \, .
\eeq
The wavefunction corresponding to a $\pi(\del_i\pi)^2$ vertex is obtained by adding up~\eqref{eq:resp0} and~\eqref{eq:resp2} with $k_*^2 \, C_{2,0}=2(1-i\a) \,C_{0,0}$:
\beq \label{eq:psipidpisq}
	\psi_{\pi(\del\pi)^2} \propto \left( \frac{k_*}{k_T} \right)^{2+i\a} (2e_2-k_T^2) \left[ (\a^2+1) e_3 + (1-i\a) k_T e_2 -k_T^3 \right] \, .
\eeq
The corresponding bispectrum is 
\beq \label{eq:bisppidpisq}
\begin{aligned}
	B_{\pi(\del\pi)^2} (k_1,k_2,k_3) \propto \frac{\sum_j k_j^2}{(k_1k_2k_3)^2\, k_T^2} & \left[ \cos\Big( \a\log (k_T/k_*) \Big) -\frac{1}{\a} \sum_{i,j} \frac{k_i}{k_j} \, \sin\Big( \a\log (k_T/k_*) \Big) \right. \\
    & \left. - \frac{1}{\a^2} \left( 2 + 2 \sum_{i\neq j} \frac{k_i}{k_j} + \frac{\sum_j k_j^3}{k_1k_2k_3} \right) \cos\Big( \a\log (k_T/k_*) \Big) \right] \, .
\end{aligned}
\eeq
This coincides with the expression computed at leading order in $\a\gg 1$ in~\cite{Chen:2010bka}, but here we provide the full (non-exponentially-suppressed) result. The symmetric squeezed limit of this bispectrum is
\beq \label{eq:slpidelpisq}
    \lim_{|\vk_l|\to 0} B_{\pi(\del\pi)^2} = - \frac{1}{\a \, k_s^3 \, k_l^3} \left[ \sin ( \a\log (2k_s/k_*) ) + \frac{3}{\a} \cos ( \a\log (2k_s/k_*) ) \right] + \mathcal{O} \left( \frac{1}{k_l} \right) \, .
\eeq

%%%%%%%%%
\subsubsection*{$\boldsymbol{p=3}$}

At three derivatives there is a new structure,
\beq \label{eq:resp3}
	\psi_3^{(3)}   = C_{0,2} \left( \frac{k_*}{k_T} \right)^{3+i\a} e_3^2 \, .
\eeq
This wavefunction coefficient corresponds to the EFToI vertex $\dot{\pi}^3$~\cite{Chen:2010bka}, and the bispectrum is
\beq \label{eq:Bp3}
    B_{\dot{\pi}^3} (k_1,k_2,k_3) = B^{(3)} (k_1,k_2,k_3) \propto \frac{1}{e_3k_T^3} \cos \Big( \a\log (k_T/k_*) \Big) \, .
\eeq
Taking the sum of~\eqref{eq:resp0}, \eqref{eq:resp2}, and~\eqref{eq:resp3} with
\beq
    k_*^3 \, C_{0,2}=- \frac{3}{2} (2+i\a) k_*^2 \, C_{2,0} = 6(2+i\a)(1+i\a) \, C_{0,0} \, ,
\eeq
gives the wavefunction coefficient coming from the EFToI vertex $\dot{\pi}(\del_i \pi)^2$, which is
\beq \label{eq:psipidotdpisq}
\begin{aligned}
	\psi_{\dot{\pi}(\del\pi)^2} \propto \left( \frac{k_*}{k_T} \right)^{3+i\a} & \Big[ 6(2+i\a)(1+i\a) e_3^2 -4(1+i\a)^2 k_Te_2e_3 \\
    & - 4(1+i\a) k_T^2e_2^2 +(11+4i\a-\a^2) k_T^3 e_3 +(i\a-3) k_T^4e_2 +k_T^6 \Big] \, .
\end{aligned}
\eeq
The corresponding bispectrum is
\beq \label{eq:bispdotpidpisq}
\begin{aligned}
	B_{\dot{\pi}(\del\pi)^2} (k_1,k_2,k_3) & \propto \frac{1}{e_3^2k_T^3} \Bigg[ \left(6e_3-4k_Te_2+k_T^3 \right) \cos \Big( \a\log (k_T/k_*) \Big) \\ 
    & - \frac{1}{\a} \cdot \frac{18e_3^2 -4k_Te_3(2e_2-k_T^2) -k_T^2e_2(4e_2-k_T^2)}{e_3} \, \sin \Big( \a\log (k_T/k_*) \Big) \vphantom{\Bigg]} \\
    & -\frac{1}{\a^2} \cdot \frac{12e_3^2-4k_Te_2e_3-4k_T^2e_2^2+11k_T^3e_3-3k_T^4e_2+k_T^6}{e_3} \, \cos \Big( \a\log (k_T/k_*) \Big) \Bigg] \, ,
\end{aligned}
\eeq
which coincides with the result computed at leading order in $1/\a$ in~\cite{Behbahani:2011it}, but here we find the other power-law-suppressed terms too. The symmetric squeezed limits of both~\eqref{eq:Bp3} and~\eqref{eq:bispdotpidpisq} satisfy:
\beq \label{eq:slp3}
    \lim_{|\vk_l|\to 0} B_{\dot{\pi}^3} \, , \, \lim_{|\vk_l|\to 0} B_{\dot{\pi}(\del\pi)^2} = \mathcal{O} \left( \frac{1}{k_l} \right) \, .
\eeq

%%%%%%%%%%%%%%%%%%%%%%%%%%%%%%%%%%%%%%%%%%%%%%%%%%%%%%%%%%%%%

\subsection{Parity-odd interactions}\label{ss:PO}

Here we briefly comment on parity-odd correlators in the context of resonant non-Gaussianity. For scalar fields, the power spectrum and bispectrum are necessary parity even and so the first correlator that can be parity-odd is the four-point function or trispectrum. In this sense, this subsection is a bit orthogonal to the main focus of this work. Our main observation is that resonant non-Gaussianity avoids the no-go theorems of \cite{Liu:2019fag,Cabass:2022rhr} and can easily induce any parity-odd $n$-point correlator at tree level for $n\geq 4$. Indeed, the oscillating coupling constants invalidate the assumptions of those no-go theorems. As a consequence, for resonant non-Gaussianity the wavefunction is generically complex, with different phases for different values of the kinematics, $\psi_n \propto k_T^{-i\alpha}$ and a parity-odd correlator picks up the non-vanishing imaginary part. As a simple example consider the EFT of single-clock inflation. Then any of the following couplings\footnote{Here $K_{\alpha\beta}$ is the extrinsic curvature of constant-clock hypersurfaces, $n^\gamma$ is the four-vector orthogonal to them, $\mathbf{e}^{\mu\nu\rho\sigma}$ is the totally anti-symmetric Levi-Civita tensor and $D^2$ is the covariant spatial Laplacian.} (see \cite{Cabass:2022rhr})
\begin{align} 
\mathbf{e}^{\mu\nu\rho\sigma}n_\mu\delta\! K_{\alpha \beta} \delta\! K^{\alpha}{}_{\nu} \nabla^{\beta} \delta\! K_{\gamma \rho} \delta\! K^{\gamma}{}_{\sigma} & \supset a^{-9} \epsilon_{ijk}\partial_m\partial_n \pi \partial_n\partial_i {\pi}\partial_m\partial_l \partial_j {\pi} \partial_l \partial_k{\pi} \,\,, \label{ZerothPOoperator} \\ 
(g^{00}+1)\mathbf{e}^{\mu\nu\rho\sigma}n_\mu\delta\! K_{\nu\lambda}(D^2\delta\! K^\lambda_{\hphantom{\lambda}\rho})\nabla_\sigma\delta\! K & \supset a^{-9}\dot{\pi}\epsilon_{ijk}\partial_i\partial_l\pi\partial_l\partial_j\partial^2{\pi}\partial_k\partial^2{\pi} \,\,, \label{FirstPOoperator} \\ 
(g^{00}+1)\mathbf{e}^{\mu\nu\rho\sigma}n_\mu\delta\! K_{\nu\lambda}(n^\alpha\nabla_\alpha\delta\! K^\lambda_{\hphantom{\lambda}\rho})\nabla_\sigma\delta\! K & \supset a^{-7}\dot{\pi}\epsilon_{ijk}\partial_i\partial_l\pi\partial_l\partial_j\dot{\pi}\partial_k\partial^2{\pi} \label{SecondPOoperator} \,\,, 
\end{align}
leads to a parity-odd trispectrum $B_4^{PO}$ already for a contact tree-level diagram. Notice that the resulting $B_4^{PO}$ has a total energy pole and hence is not in factorized form, in contrast to what one finds for constant couplings in de Sitter \cite{Stefanyszyn:2023qov}.

%%%%%%%%%%%%%%%%%%%%%%%%%%%%%%%%%%%%%%%%%%%%%%%%%%%%%%%%%%%%%

\subsection{Infrared divergences}\label{ss:IR}

So far we have neglected the possibility that wavefunction coefficients do not converge as $\eta \to 0$, a case which we will refer to as \textit{IR divergences}. These divergences are expected to appear for interactions with few derivatives and more precisely when $2n_t+n_i<4$, where $n_t$ and $n_i$ count the number of time and space derivatives, respectively. Here we discuss these terms in detail. The upshot is that there are several contributions to the wavefunction but only one contribution to the bispectrum of a spectator field, which furthermore must vanish for the bispectrum of curvature perturbations in single field inflation.\\

The main observation is that, in the presence of IR divergences, we have to choose a late time cutoff $\eta_0$ to evaluate the wavefunction coefficients. As a result, our discrete scale invariance constraint becomes
\begin{align} \label{newscaleinv}
	\exists \quad \bar{\lambda}\in (1,\infty) \qquad\text{s.t.}\qquad \psi_3(\bar{\la}^n k_a,\eta_0/\bar \la^n ) = \bar{\la}^{3n} \, \psi_3 (k_a,\eta_0) \qquad \forall \,\, n\in\mathbb{Z} \, .
\end{align}
If we furthermore impose locality in the form of manifestly local interactions, we are guided towards and ansatz containing the possible divergences $1/\eta_0^3$, $1/\eta_0^2$, $1/\eta_0^1$ and possibly a logarithmic term, each multiplied by polynomials in the norm of the momenta and possibly a function that breaks continuous scale invariance to the discrete subgroup. Because we are interested in the late time contributions, there is no need to resort to the resonance approximation and instead we proceed considering the full contribution induced by the real coupling constant $\cos \omega t$, so that we can use the cosmological optical theorem (COT) in its original form. After carefully studying the bulk integral, we arrive at the following ansatz for the IR divergent terms
\beq
\begin{aligned}\label{last}
    \psi_3(k_a,\eta_0)& \, \supset \, \frac{\text{Poly}_0(k_a) \cos(\alpha \log(-H \eta_0)+\varphi_3)}{\eta_0^3}+\frac{\text{Poly}_1(k_a) \cos(\alpha \log(-H \eta_0)+\varphi_2)}{\eta_0^2}\\
    & \quad +\frac{\text{Poly}_2(k_a) \cos(\alpha \log(-H \eta_0)+\varphi_1)}{\eta_0}+\text{Poly}_3(k_a)\cos(\alpha \log(-H\eta_0)+\varphi_0)\,,
\end{aligned}
\eeq
where $\text{Poly}_{0,1,2} (k_a)$ are homogeneous polynomials in the momenta of the indicated degree (so that $\text{Poly}_0$ is just a complex number) and $\varphi_{1,2,3}$ are real phases. This ansatz can be justified as follows. Consider the time integral of the form
\beq 
	\psi_3^{\text{contact}} (k_a) = \int_{-\infty}^0 \frac{\rd\eta}{\eta^4} \, \cos(\omega t) e^{i  k_T \eta} \, f(k_a,\eta)\,.
\eeq
Since we are interested in IR divergences, we focus on the late time contributions coming from the part of the integral at very small $\eta$. Then we can expand the exponential and the polynomial function $f$ and keep only the IR divergent terms in the integrand
\begin{align}
    \psi_3^{\text{contact}} (k_a) \supset \int \rd\eta \, \cos(\a \log(-\eta H)) \left[ \frac{\text{Poly}_0(k_a)}{\eta^4}+\frac{\text{Poly}_1(k_a)}{\eta^3}+\frac{\text{Poly}_2(k_a)}{\eta^2}+\frac{\text{Poly}_3(k_a)}{\eta}+\dots\right]\,.
\end{align}
Now the indefinite integral can be easily performed and gives precisely the terms shown in the ansatz.\\

To proceed, we notice that the COT imposes that the terms even in $k_a$ and hence odd in $\eta_0$ are purely imaginary. These terms are therefore present in the wavefunction but do not contribute to the correlator, because the latter picks up just the real part of $\psi_n$ (for parity-even interactions, which is always the case for scalar bispectra). Then we notice that the manifestly local test demands $\text{Poly}_1 =0$, because all linear terms in $k_a$ must vanish by manifest locality. The expert reader will have noticed that this is verbatim what happens in the case of scale invariant IR divergences in the wavefunction, see \cite{MLT} for a detailed discussion. Moreover, the MLT imposes that the only possible symmetric polynomial of degree three is the sum of $k_a^3$ and the overall coefficient must be real by virtue of the COT. Hence the only term that can contribute to the correlator is 
\begin{align}\label{last0}
    \psi_3(k_a,\eta_0)&\supset \sum_a k_a^3 \cos(\alpha \log(\eta_0/\eta_\ast))\,,
\end{align}
where we have re-absorbed the phase into the constant $\eta_*$. Notice that this is just the well-known and loved local non-Gaussianity, albeit with an amplitude that oscillates with time
\begin{align}\label{lastB}
B (k_a,\eta_0) \sim \cos(\alpha \log(\eta_0/\eta_\ast))  \frac{\sum_a k_a^3}{\prod_a k_a^3}  \,.
\end{align}
Several aspects of this result are novel and interesting. \\

First, notice that this local contribution is generated by all interactions with $2n_t+n_i<4$, namely $\phi^3$, $\dot \phi \phi^2$ and $(\partial_i\phi)^2\phi$. When these interactions have a coupling that is constant in time, they lead to IR divergences also in the correlator in the form of a term proportional to $\log(-k_T \eta)$. These divergences are to be expected because, as we approach the future conformal boundary of dS, the corresponding interaction acts for an infinite amount of time, in contrast to interactions with more derivatives, which redshift away. Often these divergences are interpreted as failures of perturbation theory rather than an actual physical instability, and a non-perturbative semi-classical treatment can sometimes resum these divergences into finite effects, such as for the more studied case of $\lambda \phi^4$ \cite{Starobinsky:1994bd}. Here instead the amplitude of the bispectrum never grows because $\cos (\alpha\log(-\eta H))$ remains bounded at all times. So perturbation theory remains valid but the time integral never converges (indeed $\eta = 0 $ is an essential singularity). \\

Second, what we have discussed so far is a generic spectator field and one should ask what happens to curvature perturbations $\zeta$. We know that in (attractor) single field inflation, $\zeta$ should become constant on super Hubble scales and this is incompatible with the time dependence in \eqref{last0}. Indeed we expect that to compute the bispectrum of $\zeta$ one would have to add to the inflation self interaction that generates \eqref{last0} also a second-order gauge transformation from $\phi$ to $\zeta$ that cancels the first contribution exactly. Indeed notice that the field redefinition
\begin{align}
    \phi \to \phi+\cos(\alpha \log(-H\eta))\phi^2\,,
\end{align}
generates exactly the contribution in \eqref{lastB}. This is presumably the reason why this contribution had not been discussed previously in the literature \cite{Flauger:2010ja,Behbahani:2011it}.\\

Third, notice that the local contribution in \eqref{lastB} oscillates with time, but does not oscillate with the $k_a$. This is in contrast to all the other non-Gaussian shapes we have discussed here and it is a reminder that this term does not come from a resonance, but rather from much slower and much later super Hubble evolution. However, this contribution still breaks scale invariance (dS dilatations involving $k_a$ and $\eta$ as in \eqref{newscaleinv}) to a discrete subgroup. For the remainder of this work we set this contribution aside and focus on the IR-convergent resonant shapes.

%%%%%%%%%%%%%%%%%%%%%%%%%%%%%%%%%%%%%%%%%%%%%%%%%%%%%%%%%%%%%%%%%%
\section{Corrected mode functions} \label{sec:c-}

Throughout this paper we have worked with the assumption that the scalar field mode function is the Bunch-Davies one, cf. Section~\ref{sec:fact}. Nevertheless, the background oscillations of the class of models we are considering generically lead to resonant particle production during the inflationary evolution~\cite{Flauger:2009ab}. As a result, the mode function of $\phi$ gets a negative frequency correction:
\beq \label{eq:pimfc}
	\phi_k (\eta) = \phi^+_k(\eta) + c^{(-)}_k (\eta) \, \phi^-_k(\eta) \, ,
\eeq
with $\phi^+_k(\eta)=(\phi^-_k(\eta))^*$ the Bunch-Davies solution in~\eqref{BDsol}. The function $c^{(-)}_k (\eta)$ quantifies the excitation of the negative frequency mode and was computed in detail in~\cite{Flauger:2010ja,Behbahani:2011it}; it looks roughly like a step function centered at the particle-production time $k\,\eta_{\rm p} = - \a/2$ that interpolates between $c^{(-)}_k (\eta) \sim 0$ at early times (before the resonance takes place) and a finite late-time value $c^{(-)}_k (0)$.

This correction to the mode function gives rise to contributions to the resonant bispectrum that differ from the Bunch-Davies ones---some of them were studied in~\cite{Chen:2010bka,Behbahani:2011it}. These contributions have two important features. Firstly, they are suppressed with respect to their Bunch-Davies counterparts, since the amplitude of the negative frequency mode is small:
\beq \label{eq:suppc-}
    c^{(-)}_k (\eta) \, \lesssim \, c^{(-)}_k (0) \, \sim \, \frac{b_*}{\sqrt{\a}} \ll 1 \, ,
\eeq 
with $b_*<1$ the monotonicity parameter. Secondly, they peak in the folded limit, where the sum of two momenta equals the third. The reason is that, if one substitutes the mode function of one of the interacting fields (let us choose the one with momentum $\vk_3$ for concreteness) by the negative frequency one, the bulk integrals~\eqref{eq:ints} become
\beq \label{eq:intsc-}
	\int_{-\infty}^0 \rd\eta \, e^{\pm i\omega t} e^{i  (k_1+k_2-k_3) \eta} \, c^{(-)}_{k_3} (\eta) \, f(k_a,\eta) \, .
\eeq
This integral has now a stationary phase point at $ (k_1+k_2-k_3) \eta_{\rm res} = -\a$, leading to factors of $k_1+k_2-k_3$ in the denominator of $\psi_3$. This would naively imply the presence of folded singularities, but a more careful analysis reveals that they are actually not there. Notice that in the limit $k_1+k_2-k_3\to 0$ the time at which the non-gaussianity is generated is pushed to the very far past, $\eta_{\rm res} \to -\infty$, much before the time $\eta_{\rm p}$ at which the negative frequency mode function is excited. But the function $c^{(-)}_{k_3} (\eta_{\rm res})$ at times $\eta_{\rm res} \ll \eta_{\rm p}$ vanishes, so the integral~\eqref{eq:intsc-} and hence the contribution to $\psi_3$ will be zero. In other words, the folded limit probes the bulk times before the negative frequency component was excited, and hence it cannot receive a contribution from such component. 
%the negative frequency mode function cannot contribute to the bispectrum before it is excited. 
Despite the suppression~\eqref{eq:suppc-} and the absence of singular behaviour, the analyses of~\cite{Chen:2010bka,Behbahani:2011it} revealed that the enhancement near the folded configuration makes the negative frequency mode contributions comparable to the Bunch-Davies bispectra. 

One could, in principle, try to bootstrap these negative frequency bispectrum contributions by applying an algorithm similar to the one developed in this paper, but one would find several obstacles. The most important one is that the arguments of Section~\ref{sec:fact} would no longer be valid---the function $c^{(-)}_k (\eta)$ is not rational and hence the result of the integral~\eqref{eq:intsc-} cannot be approximated by the simple ansatz~\eqref{simply}. In fact, no exact solution has been written for these time integrals and the previous literature has always worked with some approximation of the function $c^{(-)}_k (\eta)$. A possible strategy for the bootstrap would be to follow~\cite{Chen:2010bka}, which modelled the step-like time dependence of $c^{(-)}_k (\eta)$ with an error function,
\beq \label{eq:Erfc-}
    c^{(-)}_k (\eta) \sim \frac{1}{2} \left[ {\rm Erf} \left( \sqrt{\frac{\a}{2}} \left( 1 - \frac{\eta}{\eta_{\rm p}} \right) \right) + 1 \right] \, .
\eeq
After evaluating~\eqref{eq:intsc-} in the stationary phase approximation, this would lead to an overall factor of
\beq \label{eq:Erf}
    c^{(-)}_{k_3} (\eta_{\rm res}) \sim \frac{1}{2} \left[ {\rm Erf} \left( \sqrt{\frac{\a}{2}} \left( 1 - \frac{\eta_{\rm res}}{\eta_{\rm p}} \right) \right) + 1 \right] = \frac{1}{2} \left[ {\rm Erf} \left( \sqrt{\frac{\a}{2}} \left( 1 - \frac{2k_3}{k_1+k_2-k_3} \right) \right) + 1 \right] \, ,
\eeq
which accounts for the ordering of events in the bulk---the negative frequency mode excitation and the non-gaussianity production---and ensures that the behaviour described in the previous paragraph is correctly reproduced. The other function in the integrand, $f(k_a,\eta)$, would still be rational in this approximation, so the ansatz for $\psi_3$ would take the form of a rational function multiplied by the oscillatory exponential and~\eqref{eq:Erf}. The Bose symmetry requirement of Section~\ref{sec:Bose} would now be less constraining, but straightforward to implement. The unitarity and locality constraints (sections~\ref{sec:COT} and~\ref{sec:MLT} respectively), however, would require a careful analysis of the properties of $c^{(-)}_k (\eta)$.\footnote{Consider for example unitarity. It was shown in~\cite{Goodhew:2021oqg} that the corrected mode function~\eqref{eq:pimfc} satisfies hermitian analyticity if the full solution for $c^{(-)}_k (\eta)$ is used. However, the approximation~\eqref{eq:Erfc-}, needed to make the integrals attainable, spoils this property, and does not seem to be compatible with the $\a\to -\a$ mapping described in Section~\ref{sec:COT} either.} We think that this continuous need of invoking the properties of the bulk propagators and integrals partially defeats the purpose of the bootstrap, so we have limited ourselves to outlining the strategy without actually pursuing it. We hope that future developments on our understanding of cosmological correlators will allow to bootstrap these negative frequency bispectra form a more ``boundary-centered'' point of view.

%%%%%%%%%%%%%%%%%%%%%%%%%%%%%%%%%%%%%%%%%%%%%%%%%%%%%%%%%%%%%%%%%%
\section{Constraints from the CMB} \label{sec:cmb}

In this section we constrain some of the primordial bispectrum shapes bootstrapped in the previous pages using Planck 2018 temperature and polarization data. For that, we make use of the recently-developed CMB-BEST pipeline~\cite{Sohn2023cmbbest}. \\

We put constraints directly on the bispectra given by the structures that we found with up to three derivatives (cf. Subsection~\ref{sec:lowestp}). By doing this, we do not take into account the fact that in single-field inflation the leading term $\propto 1/k_l^3$ in the squeezed limit does not contribute to local observables~\cite{Pajer:2013ana,Cabass:2018roz}. Hence, if we want to interpret our bispectra as correlators of the primordial curvature perturbations $\zeta$, they can only arise in a multi-field inflationary theory. We have used the high-resolution ``Legendre'' basis of CMB-BEST, with mode number $p_{\rm max}=30$. This allows for an accurate analysis of our shapes in a reasonable range of frequencies without the need for developing a targeted resonant basis. %---we leave this possibility for a potential, more exhaustive future analysis. 
Nevertheless, the accuracy of the constraints decreases as the frequency $\alpha$ of oscillations increases, mostly due to the difficulty for the chosen basis to capture the rapid oscillations of our shapes at small $k_T$; we have found in all cases that the ``convergence epsilon'', which the code uses to estimate the offset in $f_{\rm NL}$, exceeds $\epsilon = 0.1$ when the frequencies reach $\alpha \sim 35-40$. At higher frequencies the basis we use does not accurately reproduce the resonant signal. For this reason, and since our results are valid in the $\alpha\gg 1$ regime, we restrict the analysis to the range $10\leq\alpha\leq 34$. We have maximized our results over the phase of the oscillations, showing constraints for the best-fit value of $k_*$ at each frequency $\a$.

We follow the conventions of~\cite{Planck:2019kim,Sohn2023cmbbest}, in which $f_{\rm NL}$ is defined by
\beq
    \langle\Phi (\vk_1)\Phi (\vk_2)\Phi (\vk_3)\rangle = (2\pi)^3 \, \delta^{(3)} (\vk_1+\vk_2+\vk_3) \, f_{\rm NL} \, B_{\Phi} (k_1,k_2,k_3) \, ,
\eeq
with $\Phi$ the Bardeen potential, which on superhorizon scales is related to the curvature perturbation $\zeta$ by $\Phi=3\zeta/5$. The bispectrum is normalized following
\beq
    k_{\rm p}^6 \, B_{\Phi} (k_{\rm p},k_{\rm p},k_{\rm p}) = 6A^2 \, ,
\eeq
with $k_{\rm p}=0.05 \,\,\text{Mpc}^{-1}$ the pivot scale and $A$ the amplitude of the potential power spectrum,
\beq
    P_{\Phi} (k) =Ak^{n_s-4} \, .
\eeq

% %%%%%%%%%%%%%%%%%%%%
% \subsection{Multi-field inflation}

\begin{figure}[h!]
\centering
\includegraphics[scale=0.7]{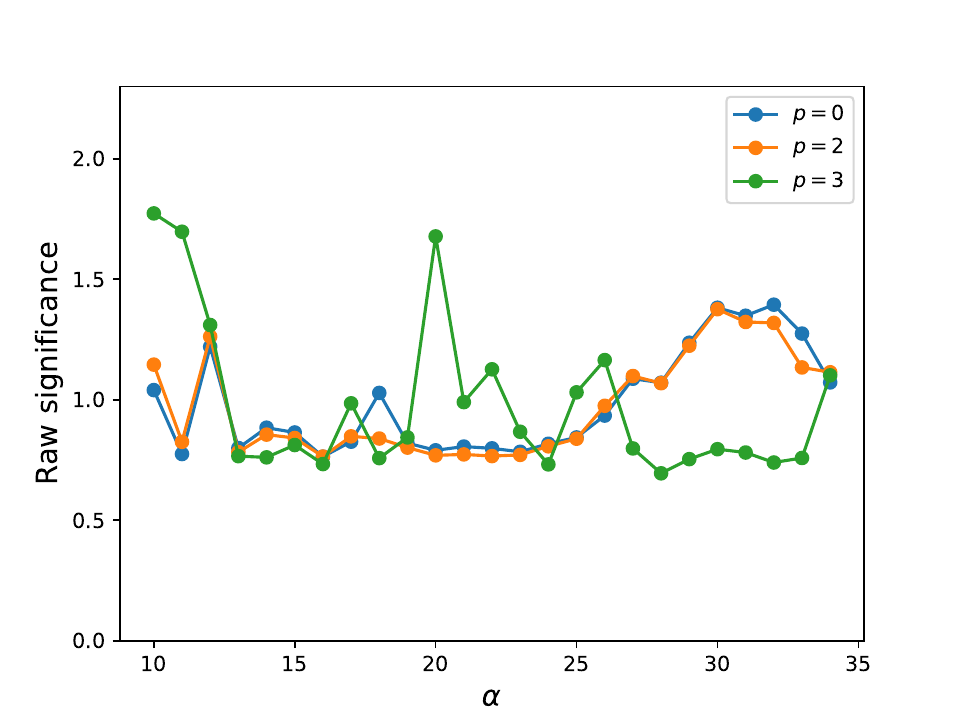}
\caption{Constraints on $f_{\rm NL}$ for the independent resonant shapes found by our bootstrap procedure at zero, two, and three derivatives, whose corresponding bispectra are given in \eqref{eq:Bp0}, \eqref{eq:Bp2}, and~\eqref{eq:Bp3} respectively. We plot the significance $|f_{\rm NL}|/\sigma (f_{\rm NL})$ as a function of the frequency $\a$ of oscillations, maximized over phase $k_*$ at each frequency. The constraints are obtained from Planck 2018 data using the CMB-BEST pipeline~\cite{Sohn2023cmbbest}.}
\label{fig:multi_indepstr}
\end{figure}

\paragraph{Results} There are only three independent structures satisfying $p\leq 3$: the shape in~\eqref{eq:resp0} at $p=0$, that in~\eqref{eq:resp2} at $p=2$, and finally the shape in~\eqref{eq:resp3} for $p=3$. %The $p=1$ structure~\eqref{eq:resp1} and the $p=3$ structure that multiplies $\widetilde{C}_{1,1}$ in~\eqref{eq:resp3} are not independent, since they just represent the lower-$p$ structures entering at subleading order in $\a$.
The results of the analysis are plotted in Figure~\ref{fig:multi_indepstr}, which shows the constraints on $f_{\rm NL}$ for the three shapes at hand. We plot the detection significance $|f_{\rm NL}|/\sigma (f_{\rm NL})$, i.e. the ratio between the absolute value of the estimated $f_{\rm NL}$ and the sample variance $\sigma (f_{\rm NL})$. We observe that for most shapes this value is below $1\sigma$, with some peaks around $1.75\sigma$. We are scanning over 25 different frequencies $\a$ for each shape, so it is natural to find some large detection values as a result of the look-elsewhere effect. A more refined analysis would require quantifying this effect and adjusting the significance accordingly, for example using the treatment in~\cite{Fergusson:2014hya}, but that goes beyond the scope of this work. The label ``raw significance'' in our plots emphasizes that we are not adjusting for the look-elsewhere effect. All things considered, we conclude that there is no significant evidence for detection of the three $p\leq 3$ resonant shapes. To a large extent this could have been anticipated given the analysis performed by the Planck collaboration \cite{Planck:2019kim}.

\begin{figure}[h!]
\centering
\includegraphics[scale=0.7]{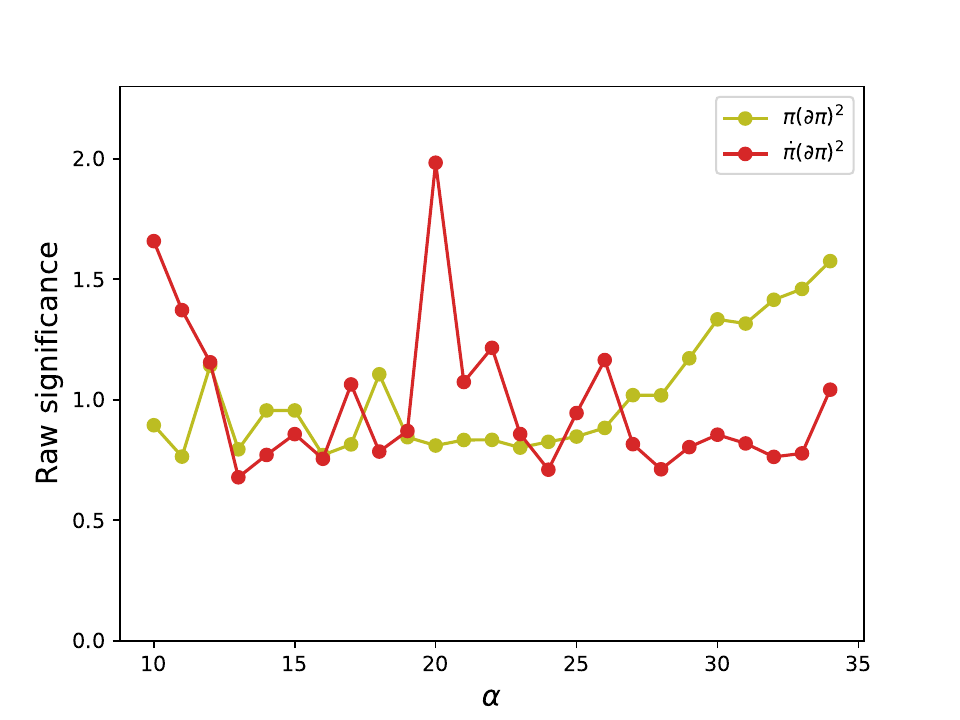}
\caption{Constraints on $f_{\rm NL}$ for the resonant shapes coming from EFToI vertices $\pi(\del\pi)^2$ and $\dot{\pi}(\del\pi)^2$, whose corresponding bispectra are given in \eqref{eq:bisppidpisq} and~\eqref{eq:bispdotpidpisq}. We plot the significance $|f_{\rm NL}|/\sigma (f_{\rm NL})$ as a function of the frequency $\a$ of oscillations, maximized over phase $k_*$ at each frequency. The constraints are obtained from Planck 2018 data using the CMB-BEST pipeline~\cite{Sohn2023cmbbest}.}
\label{fig:multi}
\end{figure}

Certain linear combinations of the structures found at $p\leq 3$ correspond to bispectra generated by vertices present in the effective field theory of inflation (EFToI ) with up to three derivatives, cf. Section~\ref{sec:lowestp}.   
There are six such vertices. For four of them, $\pi^3$, $\pi^2\dot{\pi}$, $\pi\dot{\pi}^2$, and $\dot{\pi}^3$, the bispectra are degenerate with the $p=0$, $p=2$ and $p=3$ structures that we have already constrained and whose results are plotted in Figure~\ref{fig:multi_indepstr}.\footnote{The bispectra of $\pi^3$ and $\pi^2\dot{\pi}$ are both given by~\eqref{eq:Bp0} and only differ by the choice of phase $k_*$, but this is irrelevant for the purposes of this section because we are marginalizing over $k_*$ to find the constraints. Vertices $\pi\dot{\pi}^2$ and $\dot{\pi}^3$ have bispectrum~\eqref{eq:Bp2} and~\eqref{eq:Bp3} respectively.} For the other two vertices, $\pi(\del\pi)^2$ and $\dot{\pi}(\del\pi)^2$, the bispectra are respectively~\eqref{eq:bisppidpisq} and~\eqref{eq:bispdotpidpisq}, and the bounds on $f_{\rm NL}$ are shown in Figure~\ref{fig:multi}. 
Just as before, we see low overall significance with a few peaks above $1.5\sigma$, the largest of which is found for $\dot{\pi}(\del\pi)^2$ at $\a=20$, with $\sim 2\sigma$. Taking into account the look-elsewhere effect, we do not find significant evidence for rejecting the null hypothesis $f_{\rm NL}=0$.

\paragraph{Single field inflation} We conclude with a comment on the non-Gaussian signal in single field inflation. Recall from Subsection~\ref{sec:lowestp} that the bootstrapped bispectra with $p=0$ and $p=2$ diverge as $1/k_l^3$ in the squeezed limit $|\bold{k}_l|\to 0$. References~\cite{Pajer:2013ana,Cabass:2016cgp} showed that, in single-field inflation, the projection to late-time local observables removes that leading term of the squeezed limit. A fully correct comparison of primordial reasonant non-Gaussianity with late time data would involve accurately taking into account the evolution after inflation and would probably leverage the simplification obtained by using conformal Fermi normal coordinates \cite{Dai:2015rda} as shown in \cite{Cabass:2018roz}. However, one could find a plausible estimate of the constraints on single-field inflation in a much simpler way as follows. One can remove by hand the leading terms of~\eqref{eq:Bp0}, \eqref{eq:Bp2} and~\eqref{eq:bisppidpisq} in the squeezed limits. We have done this and then compared to Planck data. We found significance below $1\sigma$ for most models, with peaks around $\sim1.7\sigma$ for a few of them. We conclude that there is also no compelling evidence for primordial non-Gaussianity of this type coming from single-field inflation.

%%%%%%%%%%%%%%%%%%%%%%%%%%%%%%%%%%%%%%%%%%%%%%%%%%%%%%%%%%%%%%%%%%
\section{Conclusions}\label{sec:conclusions}

In this work we have demonstrated how to generalize the (boostless) cosmological bootstrap beyond exact scale invariance and we have computed an infinite class of so-called resonant bispectra, characterized by periodic oscillations in the sum of the norms of the momenta. We have re-derived the resonant bispectra that were known in the literature and we have computed in closed form additional shapes corresponding to tree-level interactions with any number of derivatives. We searched Planck data for these signals finding no evidence. In addition to the phenomenological implications of our work, our results are interesting from other points of view. First, we have demonstrated that, to leading order in the large frequency of oscillations, resonant non-Gaussianity is fully specified by a flat-space amplitude \textit{for all physical configurations}, and not just close to the unphysical total energy pole. In this sense, this a concrete mechanism to ``uplift'' Minkowski amplitudes to de Sitter wavefunction coefficients without deforming them. Second, our final result can be thought of as giving a meaning to the apparently nonsensical question ``what is the bispectrum for an interaction with a complex number of derivatives?'' Indeed, we have shown that resonant non-Gaussianity can be obtained by adding an imaginary part to the integer parameter $p$ that counts the order of the total energy singularity. Third, we have demonstrated the power of general constraints such as locality in the form of the manifestly local test, by discovering an (inconsequential) inconsistency at subleading order in the results published in the literature. We have also shown how to modify the constraints of unitarity in the form of the cosmological optical theorem to leverage the analytic dependence of the result on a parameter of the model, namely the frequency of oscillations. \\

\noindent Our results can be extended in a series of promising directions:
\begin{itemize}
    \item As we discussed in Section \ref{sec:c-}, there are additional contributions coming from the excitation of negative frequency modes away from the Bunch-Davies vacuum. It would be interesting to see if these can be further constrained or even computed explicitly using bootstrap techniques.
    \item While we relaxed exact scale invariance, we still assumed a discrete form of scale invariance to land on a sufficient constrained ansatz. In so doing, we have ignored the physical effect of frequency drifting, which can be caused for example by slow-roll corrections \cite{Flauger:2014ana}. The power spectrum templates in Section 6.1 of~\cite{Flauger:2014ana} suggest that a simple deformation of our ansatz (that is still compatible with discrete scale invariance~\eqref{eq:avgsi}) might capture this drift. It would be intersting to investigate if helps to provide a better fit to the data. 
    \item We hope that this work is a first step towards extending the cosmological bootstrap into a wider range of inflationary setups. In particular, it would be interesting to apply the lessons learned here to other patterns of scale-invariance breaking, like sharp potential steps or bumps that lead to sinusoidal running of the bispectrum~\cite{Chen:2006xjb,Chen:2008wn}. Other interesting targets would be the non-Gaussianities induced by heavy particle production via a discrete-shift-symmetric coupling to the inflaton~\cite{Flauger:2016idt,Munchmeyer:2019wlh} or via features in the potential~\cite{Chen:2022vzh,Werth:2023pfl}.
\end{itemize}

\paragraph{Acknowledgements} We would like to thank Xingang Chen, Paolo Creminelli, Gerrit Farren, James Fergusson, Mehrdad Mirbabayi, Sebastien Renaux-Petel, Eva Silverstein, Wuhyun Sohn and Bowei Zhang for useful discussions. E.P. has been supported in part by the research program VIDI with Project No. 680-47-535, which is (partly) financed by the Netherlands Organisation for Scientific Research (NWO). This work has been partially supported by STFC consolidated grant ST/T000694/1 and ST/X000664/1 and by the EPSRC New Horizon grant EP/V017268/1.

\appendix

%%%%%%%%%%%%%%%%%%%%%%%%%%%%%%%%%%%%%%%%%%%%%%%%%%%%%%%%%%%%%%%%%%
\section{Analytic structure} \label{app:anstr}

In these notes we assumed the ansatz 
\beq
	\psi_3 (k_a) = e^{-i\a\log{(k_T/k_*)}} \cdot \hat{\psi}_3 (k_a) \quad\text{with $\hat{\psi}_3 (k_a)$ rational} \, ,
\eeq
as a minimal solution of the discrete scale invariance condition. We justified this in the main text by studying the bulk integrals that give rise to $\psi_3$. However, in the spirit of the bootstrap, we would like to find an alternative argument in favor of this ansatz that does not rely on bulk considerations. We sketch a plausible argument in the following, fleshing out the hand-wavy discussion given in Section \ref{sec:fact}.

The starting point is the condition~\eqref{eq:avgsi} of discrete scale invariance: our $n$-point wavefunction coefficients must satisfy
\beq \label{eq:avgsiA}
	\exists \quad \bar{\lambda}\in (1,\infty) \qquad\text{s.t.}\qquad \psi_n(\bar{\la}^mk_a) = \bar{\la}^{3m} \, \psi_n (k_a) \qquad \forall \,\, m\in\mathbb{Z} \, .
\eeq
The usual scale-invariant wavefunction coefficients (for example $\psi_2 (k) \propto k^3$ for the two-point case) satisfy this constraint for general $\bar{\lambda}$, but we will be interested in functions that break scale invariance and so are exclusive to resonant models. 

As we consider a "boundary" quantity such as the $\eta \to 0$ limit of a wavefunction coefficient, the organising principle that reflects (and hence substitutes) the bulk perturbative expansion is the analytic structure of the wavefunction. A boundary strategy then consist in imposing the simplest possible analytic structure in the hope to capture the leading tree level contributions in the bulk. For simplicity, we restrict our discussion here to the two-point wavefunction coefficient $\psi_2 (z)$, which we now take to depend on a single complex variable. This avoids the complications of multi-dimensional complex analysis that arise for higher-point objects.

%We will now start from the simplest possible analytic structure for $\psi_2 (z)$ and then progressively increase the complexity of the analytic structure until we find a function different from the scale-invariant case.

Let us now rank the possible analytic structures of $\psi_2 (z)$ in increasing level of complexity, until we find a sufficiently rich new ansatz:

\begin{enumerate}
	\item \textbf{Analytic in the whole extended $\mathbb{C}$-plane except maybe at $z=0$ and/or $z=\infty$}
	
	In this case the function $\psi_2 (z)$ can be written as a Laurent series
	\beq
		\psi_2 (z) = \sum_{n=-\infty}^{\infty} c_n z^n \, ,
	\eeq
	for $0<|z|<\infty$. The coefficients are
	\beq
		c_n = \frac{1}{2\pi i} \oint_C \frac{\psi_2(z)}{z^{n+1}} \rd z \, ,
	\eeq
	where $C$ is a positively oriented contour that encloses $z=0$. Using~\eqref{eq:avgsiA} we have, $\forall \,\, m\in\mathbb{Z}$,
	\beq
		c_n = \frac{1}{2\pi i \, \bar{\la}^{3m}} \oint_C \frac{\psi_2^{(B)}(\bar{\la}^m z)}{z^{n+1}} \rd z = \frac{\bar{\la}^{\, m (n-3)}}{2\pi i} \oint_{C_m} \frac{\psi_2^{(B)}(z_m)}{z_m^{n+1}} \rd z_m = \bar{\la}^{\, m (n-3)} \, c_n \, ,
	\eeq
	where $z_m\equiv \bar{\la}^m z$ and $C_m$ is obtained by multiplying every $z\in C$ by $\bar{\la}^m$. Hence, we see that for every $n,m\in\mathbb{Z}$ and some $\bar{\la}\in (1,\infty)$ we must have $c_n=\bar{\la}^{\, m(n-3)} c_n$. But this is only possible if $c_n=0$ for all $n\neq 3$, i.e. if the wavefunction coefficient takes the scale-invariant form $\psi_2(z)\propto z^3$ in $0<|z|<\infty$. We conclude that for this simplest analytic structure, a function $\psi_2 (z)$ different from the standard scale-invariant one does not exist.
	
	\item \textbf{Isolated singularity/ies at $z_0\neq \{ 0, \infty \}$}
	
	If an isolated singularity is present at a location $z=z_0$ with $0<|z_0|<\infty$, then by the property~\eqref{eq:avgsiA} there are also infinitely many isolated singularities at $z=\bar{\la}^m \, z_0$ for every $m\in\mathbb{Z}$, see Figure~\ref{fig:poles}. In other words, if isolated singularities are present at locations different from the origin or infinity, then there are infinitely many of them. This infinite sequence of isolated singularities is such that any arbitrarily small circle around $z=0$ contains infinitely many of them. This means that the singularities accumulate at $z=0$, which is then a cluster point.
	We don't know of any argument why such a functional form would not be consistent with~\eqref{eq:avgsiA}, but notice that it would single out a (complex but finite) momentum scale $z_0$ (or several of them if we consider isolated singularities at several points not related by an integer power of $\bar{\lambda}$). We will then dismiss this analytic structure as unphysical, since neither the setup of the problem nor the symmetry property~\eqref{eq:avgsiA} select a particular scale.
	%Since there is no Laurent series representation valid in the neighbourhood of a cluster point, I don't think we can use Cauchy's integral formula to rule out the existence of these functions.
	
	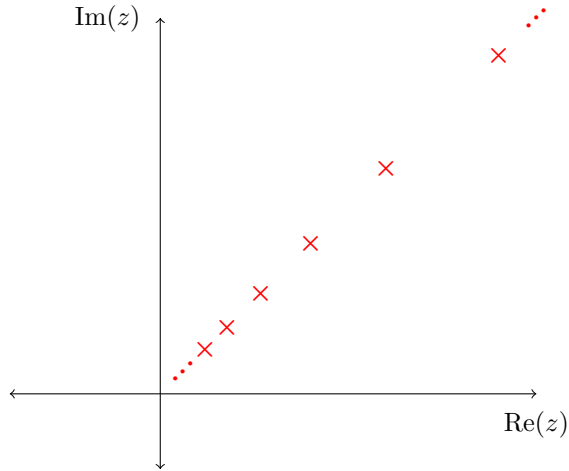
\begin{figure} 
	\centering
	\begin{tikzpicture}
	
	\def\L{1.5}
	\def\Z{2}
	
	\draw[<->] (-2,0) -- (5,0);
	\draw[<->] (0,-1) -- (0,5);
	\node() at (5,-0.4) {Re$(z)$};
	\node() at (-0.7,5) {Im$(z)$};
	
% 	\node[red](O) at (0,0) {$\boldsymbol{\otimes}$};
	
%	\node[red](A-4) at (\Z/\L^4,\Z/\L^4) {$\boldsymbol{\times}$};
	\node[red](A-3) at (\Z/\L^3,\Z/\L^3) {$\boldsymbol{\times}$};
	\node[red](A-2) at (\Z/\L^2,\Z/\L^2) {$\boldsymbol{\times}$};
	\node[red](A-1) at (\Z/\L,\Z/\L) {$\boldsymbol{\times}$};
	\node[red](A0) at (\Z,\Z) {$\boldsymbol{\times}$};
	\node[red](A1) at (\Z*\L,\Z*\L) {$\boldsymbol{\times}$};
	\node[red](A2) at (\Z*\L*\L,\Z*\L*\L) {$\boldsymbol{\times}$};
	
	\node[red](B1) at (\Z*\L*\L+0.4,\Z*\L*\L+0.4) {$\boldsymbol{\cdot}$};
	\node[red](B2) at (\Z*\L*\L+0.5,\Z*\L*\L+0.5) {$\boldsymbol{\cdot}$};
	\node[red](B3) at (\Z*\L*\L+0.6,\Z*\L*\L+0.6) {$\boldsymbol{\cdot}$};
	
	\node[red](C1) at (0.2,0.2) {$\boldsymbol{\cdot}$};
	\node[red](C2) at (0.3,0.3) {$\boldsymbol{\cdot}$};
	\node[red](C3) at (0.4,0.4) {$\boldsymbol{\cdot}$};
	
	\end{tikzpicture}
	\caption{Possible analytic structure of $\psi_2(z)$ in the $\mathbb{C}$-plane. An isolated singularity at $z_0\neq\{ 0,\infty\}$ repeats itself at $z=\bar{\la}^m z_0$ for every $m\in\mathbb{Z}$. This means that the infinite sequence of singularities extends towards $z=\infty$ and accumulates towards $z=0$, which becomes a cluster point.}
	\label{fig:poles}
	\end{figure}

	\item \textbf{Branch points at $z=0$ and $z=\infty$}
	
	The next level of complexity of the analytic structure comes when we consider multivalued functions and the associated branch points. If a function is multivalued, it must have at least two branch points in the extended complex plane $\bar{\mathbb{C}} \equiv \mathbb{C} \cup \{\infty\}$. By the previous argument, if we had a branch point at some finite $0<z<\infty$ then we must also have its infinitely many images. The only possibility with a finite number of branch points is hence a function with two branch points at $\{ 0, \infty \}$ and a branch cut connecting them. A rich class of functions with this structure take the form $e^{-i\a\log{(z)}} \cdot \hat{\psi}_2 (z)$ where $\hat{\psi}_2 (z)$ is a homogeneous function of degree three, $\hat{\psi}_2 (\lambda z)=\lambda^3\hat{\psi}_2 (z)$. Extending it to the multi-variable three-point case $\psi_3 (k_a)$, this is the ansatz we study in this work.
\end{enumerate}

%%%%%%%%%%%%%%%%%%%%%%%%%%%%%%%%%%%%%%%%%%%%%%%%
\bibliographystyle{JHEP}
\bibliography{refs}
\end{document}